\documentclass[a4paper,preprintnumbers,floatfix,superscriptaddress,pra,twocolumn,bibnotes,showpacs,notitlepage,longbibliography]{revtex4-2}
\usepackage[utf8]{inputenc}
\usepackage[T1]{fontenc}
\usepackage[colorlinks,breaklinks,linkcolor=magenta,citecolor=blue,urlcolor=blue]{hyperref}
\usepackage{amsmath,amssymb,amsthm,bm,mathtools,amsfonts,mathrsfs,bbm,dsfont}
\usepackage{physics}
\usepackage[caption=false]{subfig}
\usepackage{graphicx}
\usepackage{xcolor}
\usepackage{float}
\usepackage{appendix}

\renewcommand{\vec}[1]{\boldsymbol{#1}}
\newcommand{\id}{\ensuremath{\mathds{1}}}
\renewcommand{\d}{\ensuremath{\mathrm{d}}}
\DeclareMathOperator*{\argmin}{arg\,min}

\begin{document}
\title{Hamiltonian and Liouvillian learning in weakly-dissipative quantum many-body systems}

\author{Tobias Olsacher}
\thanks{These two authors contributed equally.} 
\affiliation{Institute for Theoretical Physics, University of Innsbruck, Technikerstraße 21A, 6020 Innsbruck, Austria}
\affiliation{Institute for Quantum Optics and Quantum Information of the Austrian Academy of Sciences, Innsbruck, Austria}

\author{Tristan Kraft}
\thanks{These two authors contributed equally.} 
\affiliation{Institute for Theoretical Physics, University of Innsbruck, Technikerstraße 21A, 6020 Innsbruck, Austria}
\affiliation{Department of Physics, QAA, Technical University of Munich, James-Franck-Str. 1, D-85748 Garching, Germany}

\author{Christian Kokail}
\affiliation{Institute for Theoretical Physics, University of Innsbruck, Technikerstraße 21A, 6020 Innsbruck, Austria}
\affiliation{Institute for Quantum Optics and Quantum Information of the Austrian Academy of Sciences, Innsbruck, Austria}
\affiliation{ITAMP, Harvard-Smithsonian Center for Astrophysics, Cambridge, MA 02138, USA}
\affiliation{Department of Physics, Harvard University, Cambridge, MA 02138, USA}

\author{Barbara Kraus}
\affiliation{Department of Physics, QAA, Technical University of Munich, James-Franck-Str. 1, D-85748 Garching, Germany}
\affiliation{Institute for Theoretical Physics, University of Innsbruck, Technikerstraße 21A, 6020 Innsbruck, Austria}

\author{Peter Zoller}
\affiliation{Institute for Theoretical Physics, University of Innsbruck, Technikerstraße 21A, 6020 Innsbruck, Austria}
\affiliation{Institute for Quantum Optics and Quantum Information of the Austrian Academy of Sciences, Innsbruck, Austria}

\date{\today}
\begin{abstract}
We discuss Hamiltonian and Liouvillian learning for analog quantum simulation from non-equilibrium quench dynamics in the limit of weakly dissipative many-body systems. We present and compare various methods and strategies to learn the operator content of the Hamiltonian and the Lindblad operators of the Liouvillian. We compare different ansätze based on an experimentally accessible ‘learning error’ which we consider as a function of the number of runs of the experiment. Initially, the learning error decreases with the inverse square root of the number of runs, as the error in the reconstructed parameters is dominated by shot noise. Eventually the learning error remains constant, allowing us to recognize missing ansatz terms. A central aspect of our approaches is to (re-)parametrize ansätze by introducing and varying the dependencies between parameters. This allows us to identify the relevant parameters of the system, thereby reducing the complexity of the learning task.  Importantly, this (re-)parametrization relies solely on classical post-processing, which is compelling given the finite amount of data available from experiments. We illustrate and compare our methods with two experimentally relevant spin models.
\end{abstract}

\maketitle

\section{Introduction}
Controllable quantum many-body systems, when scaled to a large number of particles, hold the potential to function as quantum computers or quantum simulators, addressing computational problems that are considered intractable for classical computers~\cite{Altman2021}. Remarkable progress has been reported recently in building quantum simulators, as programmable special-purpose quantum devices, to solve quantum many-body problems efficiently, which finds applications in condensed matter~\cite{Tarruell2018}, high-energy physics~\cite{Bauer2023}, and quantum chemistry~\cite{McArdle2020}, in both equilibrium and non-equilibrium dynamics. Quantum simulation can be realized as analog or digital quantum simulators. In analog simulation, a target Hamiltonian finds a natural implementation on a quantum device, exemplified by ultracold bosonic and fermionic atoms in optical lattices as Hubbard models~\cite{GrossBloch2017,Sompet2022,Leonard2023,Zhang2023}, or spin models with trapped ions~\cite{Monroe2021,Joshi2023,Kiesenhofer2023,Guo2024}, Rydberg tweezer arrays~\cite{Weimer2010,Semeghini2021,Ebadi2021,Chen2023}, and superconducting qubits~\cite{Kim2023,Tindall2024,Rosenberg2024}. The unique feature of analog quantum simulators is the scalability to large particle numbers. In contrast, digital quantum simulation~\cite{Pastori2022} represents the time evolution of a given many-body Hamiltonian using a freely programmable sequence of Trotter steps implemented via single and multi-qubit entangling quantum gates.

An outstanding challenge in quantum simulation is the ability to predict properties of many-body observables with controlled error while scaling to a regime of potential quantum advantage~\cite{Trivedi2022,Kashyap2024,Cai2023,Daley2022}. Given the increase of complexity of these systems, methods to characterize, and thus verify, the proper functioning of quantum simulators are required~\cite{Hangleiter2017,Eisert2020,Carrasco2021}. This includes verification, that the correct many-body Hamiltonians are being implemented, and a complete characterization of (weak) decoherence due to unwanted couplings to an environment or fluctuating external fields. In the present paper, we approach this goal by studying Hamiltonian and Liouvillian learning for analog quantum simulators. Previous works have studied various scenarios for Hamiltonian and Liouvillian learning~\cite{Wiebe2014,Holzapfel2015,Wang2015,Evans2019,Bairey2019,Bairey2020,Gu2024,Haah2024,Li2020,Gu2024,Haah2024,Ott2024,Zubida2021,França2022,Hangleiter2021}, for instance, by comparing with a trusted simulator~\cite{Wiebe2014}, or based on the preparation of steady states~\cite{Evans2019,Bairey2019,Bairey2020,Gu2024,Haah2024}. Alternative approaches are based on dynamics in (long-time) quenches~\cite{Holzapfel2015,Li2020,Gu2024,Haah2024,Ott2024}, the estimation of the time-derivatives of few-qubit observables from short-time evolution~\cite{Zubida2021,França2022}, or require additional resources such as intermediate gates~\cite{Huang2023,Yu2023}.

Here, we will be interested in Hamiltonian and Liouvillian learning from dynamics in long-time quenches, with only few experimental requirements such as the preparation of product states, and product measurements. We assume that the dynamics of the experimental quantum simulator is described by a master equation, where the Hamiltonian acts as a generator of the coherent many-body dynamics, while Lindbladian terms model the noise. The goal of Hamiltonian and Liouvillian learning is to learn the \emph{operator structure}, reminiscent of a \emph{principal component analysis}~\cite{Mehta2019}, of both the many-body Hamiltonian, as one-, two- or few-body interaction terms including their couplings, and the quantum jump operators in the dissipative Liouvillian, representing local or non-local (global) quantum and classical noise. The scalability and efficiency of Hamiltonian and Liouvillian learning are related to the assumption that physical Hamiltonians and Liouvillians will only involve few-body interactions and quantum jump operators, leading to a polynomial scaling of the number of terms to be learned with system size.

Our study below discusses and compares various scenarios and strategies of Hamiltonian and Liouvillian learning, which we illustrate by simulating learning protocols for various model cases. Our work is motivated by present trapped-ion experiments, where quantum simulators realize 1D interacting weakly dissipative spin-1/2 chains. This allows quench experiments to be performed, and we will be interested in learning the Hamiltonian and Liouvillian from experimental quench data observed at various quench times. Learning the Hamiltonian and Liouvillian requires many experimental runs. In each run projective measurements of spins are performed in various bases, allowing to measure multi-spin correlation functions up to shot noise. In addition, learning protocols will prepare many initial states, which in our case can be pure or mixed, thus resulting in stability against state-preparation errors. A central aspect of our study below will thus be an investigation of the experimentally measurable learning error of Hamiltonian and Liouvillian, and its scaling with the number of measurement runs.

The paper is structured as follows. 
Section~\ref{sec:motivation} outlines the specific scenario we're examining and the core theoretical concepts involved. Section~\ref{subsec:background} establishes some constraints on the system’s Hamiltonian and Liouvillian, measurable through simple quench experiments. We then present the main equations that enable us to infer the Hamiltonian and Liouvillian from experimental data in Sections~\ref{subsec:LearningEq} and~\ref{sec:energyConservation}. The first approach is based on the equation of motion of the expectation value of general observables, known as Ehrenfest's theorem, while the second approach is based on generalized energy conservation. In Section~\ref{sec:examples}, we compare the two methods and showcase learning protocols for various model scenarios through numerical simulations. We compare different ansätze for the operator content, using an experimentally measurable quantity, which we identify as a learning error. These ansätze are derived by re-parametrizing our ansatz, typically involving data recycling and classical post-processing. The learning process can be divided into two phases as a function of experimental runs $N_{\rm runs}$: in the early phase, the learning error is dominated by shot-noise and decays $\sim N_{\rm runs}^{-1/2}$. In the later phase, systematic errors become dominant due to missing terms in the ansatz, or an insufficient ansatz. This leads to a constant learning error independent of the number of measurements, indicating the need to extend our ansatz.

\section{Hamiltonian and Liouvillian Learning of Many-Body Systems}\label{sec:motivation}
\subsection{Background}
\label{subsec:background}

\begin{figure*}
    \centering
    \includegraphics[width=0.9\linewidth]{figures/HL.pdf}
    \caption{A quantum simulation experiment starts with a specific target Hamiltonian $H^{\rm sim}$ of interest, for instance, spin-models that we consider in this work. This Hamiltonian finds a natural realization in a quantum simulation platform that aims to simulate the dynamics of $H^{\rm sim}$ in a well controlled environment. In practice, however, the precise Hamiltonian such a device is implementing is unknown and needs to be determined. In addition, the system is typically (weakly) coupled to an environment so that the the dynamics is well described by a quantum master equation, cf. Eq.~\eqref{eq:lindblad}. Hamiltonian and Liouvillian learning aims to reconstruct the Hamiltonian and Liouvillian using a finite measurement budget. Here, we consider simple quench experiments with product states as inputs, and product measurements. Relevant questions that need to be answered include, but are not limited to, the existence of additional (small) terms in the Hamiltonian, as well as the operator content of the Liouvillian, i.e., the quantum jump operators and dissipation rates. This information can then be used as a feedback to the experiment, improving the quantum simulation, or to verify that indeed the correct dynamics has been implemented.
    }
    \label{fig:HL}
\end{figure*}

We consider analog quantum simulation in a regime, where the engineered quantum many-body system of interest is weakly coupled to a decohering environment. We assume that the system dynamics is described by a master equation with Lindblad form~\cite{gardiner2004quantum},
\begin{eqnarray}\label{eq:lindblad}
    \frac{\d}{\d t}\varrho &=& 
    -i [H,\varrho] + \frac{1}{2} \sum_k \gamma_k \left( [l_k \varrho, l_k^\dagger] + [l_k, \varrho l_k^\dagger] \right) \\ &\equiv& \mathcal{H}(\varrho) + \mathcal{L}(\varrho), \nonumber
\end{eqnarray}
comprising a coherent term $\mathcal{H}(\varrho)=-i[H,\varrho]$ with many-body \emph{Hamiltonian}  $H$ and a dissipative term $\mathcal{L}$, referred to as the \emph{Liouvillian}. The Lindblad quantum jump operators $l_k$ describe dissipative processes coupling the system to an environment. Here, $\gamma_k\geq 0$ is the physical domain for the corresponding damping rates, where the dynamics is described by a completely positive and trace-preserving map~\cite{Gorini1976,Note0}\footnotetext[0]{The Hamiltonian, $H$, the dissipation rates, $\gamma_k$, and the Lindblad operators, $l_k$, are uniquely determined by the dynamics, if one requires the following: (i) $H$ is traceless, (ii) the $l_k$ are traceless and orthonormal, i.e., $\mathrm{tr}(l_k)=0$ and $\mathrm{tr}(l_m^\dagger l_n)=\delta_{mn}$, and (iii) the $\gamma_k$ are not degenerate~\cite{Gorini1976}.}. Throughout this work, we assume that the Hamiltonian and the Liouvillian are time-independent. In typical experimental settings, the Hamiltonian terms are significantly larger than dissipative processes $ \norm{H}\gg \gamma_k$.

Below we will be interested in analog quantum simulation of spin-1/2 systems, as implemented with trapped ions~\cite{Monroe2021,Joshi2023,Kiesenhofer2023,Guo2024}, Rydberg tweezer arrays~\cite{Semeghini2021,Ebadi2021,Chen2023}, or superconducting circuits~\cite{Kim2023,Rosenberg2024}. An illustrative example of a two-local spin-Hamiltonian in one spatial dimension, which we have in mind, is given by
\begin{eqnarray} \label{eq:model1H}
    H &=& \sum_{i=1}^{N-1}  J_{i,i+1}^z \sigma_i^z \sigma_{i+1}^z + \sum_{i=1}^{N-2} J_{i,i+2}^z \sigma_i^z \sigma_{i+2}^z \\ && \quad + B_x \sum_{i=1}^N \sigma_i^x + B_z \sum_{i=1}^N \sigma_i^z, \nonumber
\end{eqnarray}
which describes a next-nearest-neighbor Ising model of $N$ spins with longitudinal- and transverse fields.

The Liouvillian, $\mathcal{L}$, in Eq.~\eqref{eq:lindblad} is defined by its Lindblad, or quantum jump operators $l_k$. Examples of Lindblad operators that typically appear in experiments include spontaneous emission, described by local Lindblad operators $l_k=\sigma^-_k$, or local dephasing, represented by $l_k=\sigma^z_k$. Besides local dissipation, we will be interested in identifying the presence of collective dissipative effects, for instance, in the form of collective dephasing caused by globally fluctuating laser fields or effective magnetic fields, leading to a collective Lindblad operator $l=\sum_{k}\sigma^z_k$.

Before we proceed with the discussion, we need to establish certain conditions on the Hamiltonian and Liouvillian of the system. For a time-independent observable, $O$, Ehrenfest's theorem, in the context of Eq.~\eqref{eq:lindblad}, states that
\begin{equation}\label{eq:ehrenfest}
    \frac{\d}{\d t}\expval*{O}= \expval{-i[O,H]}+\frac{1}{2}\sum_k\gamma_k \langle l_k^\dagger [O, l_k] + [l_k^\dagger, O] l_k \rangle,
\end{equation}
where $\expval*{X}=\tr[X\varrho]$, for any state $\varrho$. In its integral form, this generalized Ehrenfest theorem yields the following condition for the Hamiltonian and Liouvillian
\begin{subequations}
\begin{multline}\label{eq:observable}
    \expval{O}_T-\expval{O}_0=-i\int_0^T \expval{\qty[O,H]} \d t \\ + \frac{1}{2}\sum_k \gamma_k \int_0^T \expval*{l_k^\dagger[O,l_k]+[l_k^\dagger,O]l_k}_{t} \d t,
\end{multline}
where $\expval{X}_t=\tr[X\exp{t(\mathcal{H}+\mathcal{L})}\varrho(0)]$. 
This condition simplifies if we consider $O=H$, which leads to 
\begin{multline}\label{eq:energy}
\expval{H}_{T}-\expval{H}_{0}= \\ \frac{1}{2}\sum_k \gamma_k \int_0^T \expval*{l_k^\dagger[H,l_k]+[l_k^\dagger,H]l_k}_{t} \d t.
\end{multline}
\end{subequations}
It describes generalized energy conservation, including the loss of total energy of the system during a quench of duration $T$. If $\gamma_k=0$ for all $k$, this equation indicates the conservation of energy. In our Hamiltonian and Liouvillian learning protocol, Eqs.~\eqref{eq:observable} and \eqref{eq:energy} will play a crucial role. We will generalize the protocol of Ref.~\cite{Li2020} for Hamiltonian learning in the absence of dissipation, and present it in a form particularly suited for learning $H$ and $\cal L$ in the limit of weak dissipation.

As a final remark, let us elaborate on the experimental procedure that we are considering which can be used to probe the conditions in Eqs.~\eqref{eq:observable} and~\eqref{eq:energy} (see also Fig.~\ref{fig:HL} for an illustration). Starting from a product state, $\ket{\psi}=\ket{\psi_1}\otimes \dots\otimes \ket{\psi_N}$, which can be experimentally prepared with high fidelity, one evolves the state under the Hamiltonian and Liouvillian for some time $t$. The resulting state, $\varrho_t=\exp[t(\mathcal{H}+\mathcal{L})]\varrho_0$, is measured in a product basis, for instance, in the Pauli basis. Clearly, the  limiting quantity, here, and in the following, will be the total number of runs of quench experiments, which we will denote by $N_{\rm runs}$. Given, however, that many-body Hamiltonians typically consist of a few quasi-local operators, many of the required measurements can be carried out simultaneously in a single run and using classical post-processing, or via the randomized measurement toolbox~\cite{Elben2022}.

\subsection{Hamiltonian and Liouvillian learning from the generalized Ehrenfest theorem}\label{subsec:LearningEq}

Often, in an experimental setting, the detailed structure of the Hamiltonian and the Liouvillian are unknown. We present here a method to learn the operator content of $H$ and $\mathcal{L}$, and the corresponding parameters from experimental data. For instance, in the context of the spin system in and below Eq.~\eqref{eq:model1H}, identifying the operator content means identifying the Pauli operators that appear in the decomposition of $H$ and the Lindblad operators $l_k$.

We start by choosing an ansatz for the operator content for the Hamiltonian and Liouvillian. Specifically, as an ansatz for $H$ we choose
\begin{equation}
    A(\vec{c})=\sum_{j=1}^{n} c_j h_j,
\end{equation}
with parameters $\vec{c}=(c_1,\dots, c_n)$, and $h_j$ traceless and hermitian for all $j$. As an example, one could choose the $h_j$ to be few-body Pauli operators. As an ansatz for the Liouvillian we choose
\begin{equation}
\mathcal{D}(\vec{d})=\frac{1}{2} \sum_{k=1}^m d_k \left( [a_k \varrho, a_k^\dagger] + [a_k, \varrho a_k^\dagger] \right),
\end{equation}
with Lindblad operators $\{a_k\}$, and parameterized by the corresponding non-negative dissipation rates, $\vec{d}=(d_1, d_2, \ldots,d_m)$. Here, $a_k$ is intended to be an ansatz operator for a single Lindblad operator, $l_k$. However, this must not be the case as we will discuss in Section~\ref{sec:energyConservation}.

Consider the equation of motion of a general observable, $O$ (not necessarily commuting with $H$), given by the generalized Ehrenfest theorem in integral form in Eq.~\eqref{eq:observable}. Inserting the ansatz, $A(\vec{c})$ for $H$, and $\mathcal{D}(\vec{d})$ for the Liouvillian, and imposing the resulting constraint for a set of observables $\{O_i\}$, with $i=1,\dots,p$, one obtains a set of linear equations for $\vec{c}$ and $\vec{d}$. These can be written as a simple matrix equation~\footnote{To be more precise, we insert the adjoint $\mathcal{D}^\dagger$ of the ansatz for the Liouvillian into Eq.~\eqref{eq:LearningEqEhrenfest}. The adjoint is defined by $\tr[X\mathcal{D}(Y)]=\tr[\mathcal{D}^\dagger(X)Y]$ for all test operators $X,Y$. We note, that both superoperators contain the same Lindblad operators and dissipation rates.};
\begin{equation}
\label{eq:LearningEqEhrenfest}
K_H \boldsymbol{c}+K_D\vec{d} = \boldsymbol{b}.
\end{equation}
Here, $K_H$ is a $p\times n$ matrices, and $K_D$ is a $p\times m$ matrix, with entries defined by 
\begin{equation}\label{eq:constraintMatrix1}
    (K_H)_{ij} = -i \int_0^{T} \expval*{ [O_i,h_j] }_{t} \mathrm{d}t,
\end{equation}
and
\begin{equation}\label{eq:constraintMatrix2}
    (K_D)_{ik} = \frac{1}{2} \int_0^{T} \expval*{a_k^\dagger [O_i, a_k] + \mathrm{h.c.}}_{t} \, \mathrm{d}t,
\end{equation}
and the vector $\vec{b}$ is defined by $b_i = \expval*{O_i}_{T} - \expval*{O_i}_{0}$. As these equations hold for any density matrix, $\varrho$, our protocol, similar to the Hamiltonian learning protocol in Ref.~\cite{Li2020}, is resistant to state-preparation errors. Moreover, although we have imposed the above equations for multiple observables, $\{O_i\}$, and a single input state, $\varrho$, one may also consider multiple input states. Note also, that similar constraints have also been used in Ref.~\cite{Bairey2020} to learn Liouvillians from their steady-states.

In the discussion above, we have implicitly assumed that our ansatz is chosen such as to contain the Hamiltonian and Liouvillian, i.e., there exist vectors $\vec{c}^H$, and $\vec{d}^\gamma$, such that $A(\vec{c}^H)=H$, and $\mathcal{D}(\vec{d}^\gamma)=\mathcal{L}$. However, in practice we do not expect this to be the case, and we will call such an ansatz \emph{ insufficient}~\footnote{As an example, the Hamiltonian may not only contain operators $h_j$, but also additional operators $h_j'$ not contained in our ansatz. On the level of Eq.~\eqref{eq:learningEq} this would result in a truncation of the matrix $M$ which then contains fewer columns than would be required in order to reconstruct $\vec{c}^H$}. Nevertheless, we can determine the parameters of our ansatz minimizing the violation of Eq.~\eqref{eq:LearningEqEhrenfest} by minimizing the squared residuals;
\begin{equation}\label{eq:learning_cost_Ehrenfest}
(\vec{c}^{\rm rec},\vec{d}^{\rm rec})=\argmin_{ (\vec{c},\vec{d}):\vec{d}\geq 0} \norm{K_H \boldsymbol{c}+K_D\vec{d} -\boldsymbol{b}}, 
\end{equation}
where $\Vert \boldsymbol{x} \Vert$ denotes the $2$-norm. In case of an insufficient ansatz, the value of the linear optimization problem, $\min_{(\vec{c},\vec{d})} \norm{K_H \boldsymbol{c}+K_D\vec{d} - \boldsymbol{b}}$, will be strictly bounded away from zero.

\begin{figure*}
    \centering
    \includegraphics[width=0.85\linewidth]{figures/HL2.pdf}
    \caption{Hamiltonian and Liouvillian learning begins by choosing an ansatz for the operator structure of the Hamiltonian and Liouvillian. Then, one measures time-resolved expectation values of correlation functions, and estimates of their integrals over a period of total time $T$. Here, we choose to measure the correlation functions at discrete points and estimate the integral using Simpson's rule. The learning progress is assessed by considering the ratio $\lambda_1/\lambda_2$, which can be directly obtained from experimental data, as a function of $N_{\rm runs}$. This allows us to compare different ansätze, and to notice when terms in the ansatz are missing. changing the ansatz involves reparametrization, which typically can be done by classical post-processing.
    }
    \label{fig:HL2}
\end{figure*}

\subsection{Hamiltonian and Liouvillian learning from generalized energy conservation}\label{sec:energyConservation}

Instead of inserting our ansatz into Eq.~\eqref{eq:observable}, we may also insert the ansatz into the much simpler generalized energy conservation condition in Eq.~\eqref{eq:energy}. Imposing the resulting constraint for a \emph{set} of initial (product) states $\{\varrho_i\}$, where $i=1,\dots,p$, leads to the following simple matrix equation;
\begin{equation}\label{eq:learningEq}
    [M_H+M_D(\vec{d})]\vec{c}=0.
\end{equation}
The matrices $M_H$ and $M_D(\vec{d})$ are $p\times n$ matrices defined by 
\begin{equation}\label{eq:constraintMatrix3}
(M_H)_{ij}=\expval*{h_j}_{i,0}-\expval*{h_j}_{i,T},
\end{equation}
and $M_D(\vec{d})=\frac{1}{2}\sum_k d_k M^{(k)}$, with
\begin{equation}\label{eq:timeTraces}
(M^{(k)})_{ij}=\int_0^T \expval*{a_k^\dagger[h_j,a_k]+[a_k^\dagger,h_j]a_k}_{i,t}\d t,
\end{equation}
and $\expval{X}_{i,t}=\tr[X\varrho_i(t)]$. In contrast to Eq.~\eqref{eq:LearningEqEhrenfest}, where we can impose constraints for multiple input stats and multiple observables, the observables appearing in Eq.~\eqref{eq:learningEq} are fixed by the ansatz. Therefore to obtain more constraints, one needs to consider more input states. Moreover, as these equations hold for any state, $\varrho$, the reconstruction is resistant to state preparation errors.

Similar as above, in the case of an insufficient ansatz that cannot fulfill Eq.~\eqref{eq:learningEq}, one can reconstruct those parameters minimizing the violation of generalized energy conservation;
\begin{equation}\label{eq:learning_cost}
(\vec{c}^{\rm rec},\vec{d}^{\rm rec})=\argmin_{ (\vec{c},\vec{d}):\norm{\vec{c}}=1,\vec{d}\geq 0} \norm{\qty[M_H+M_D(\vec{d})]\vec{c}}.
\end{equation}
Without dissipation, where $\vec{d}=0$, this is a linear optimization problem, while in the presence of dissipation, i.e., $\vec{d}>0$, Eq.~\eqref{eq:learning_cost} involves non-linear optimization over $\vec{c}$ and $\vec{d}$. Note, that for a fixed $\vec{d}$, the minimization over $\vec{c}$ results in the smallest singular value, $\lambda_1$, of the constraint matrix $M_H+M_D(\vec{d})$. Computing the minimum of the cost function in Eq.~\eqref{eq:learning_cost} then corresponds to finding non-negative dissipation rates $\vec{d}^{\rm rec}$ for which the smallest singular value, $\lambda_1$, of the constraint matrix $M_H+M_D(\vec{d})$ attains its minimum. Numerically we find $\vec{d}^{\rm rec}$ using the "Dividing Rectangles" (DIRECT) algorithm implemented in SciPy with default settings~\footnote{The lower bound for each $d_k$ is set to $d_k^\mathrm{min}=0$, and the upper bound is set proportional to the maximum dissipation rate of the model $d_k^\mathrm{max}= k \gamma_\mathrm{max} = k \mathrm{max}(\{ \gamma_k\})$. We check convergence for different values of $k$ and different numbers of cost-function evaluations.}. The corresponding $\vec{c}^{\rm rec}$ is then the right-singular vector corresponding to $\lambda_1$. In the following we will denote the singular values of the constraint matrix by $\lambda_n\geq\dots\geq\lambda_2\geq\lambda_1\geq 0$.

\subsubsection{Learning the overall scale, and conserved quantities}

The generalized energy conservation conditions in Eq.~\eqref{eq:energy} define the Hamiltonian only up to a scalar factor since they are linear in $H$. Additionally, these conditions hold for any operator $Q$ with $[Q,H]=0$, i.e., for all \emph{conserved quantities} of $H$. An ansatz, $A(\vec{c})$, then may contain the Hamiltonian, $H$, and other conserved quantities of $H$ which admit a decomposition in the form of $A(\vec{c})$. For instance, if we choose our ansatz to be $k$-local, it contains at most $k$-local conserved quantities, such as the total magnetization, $Q\sim \sum_k\sigma^z_k$, which is a sum of local operators.

So let us assume that an ansatz contains two linearly independent conserved quantities $H$, and $Q$, corresponding to two linearly independent vectors $\vec{c}^H$, and $\vec{c}^Q$ (the generalization is straightforward). Then, by singular value decomposition, one obtains two right singular vectors $\{\vec{v},\vec{v}_\perp\}$, which both satisfy Eq.~\eqref{eq:learningEq}, and degenerate singular values $\lambda_1=\lambda_2=0$. However, the right singular vectors must not necessarily correspond to $\vec{c}^H$ and $\vec{c}^H_\perp$. Thus, in the presence of conserved quantities, naively solving Eq.~\eqref{eq:learning_cost} would in general reconstruct the parameters of a linear combination of conserved quantities.

To distinguish $\vec{c}^H$ from other conserved quantities, we impose additional constraints that can only be fulfilled by $H$, but not by other conserved quantities, including scalar multiples of $H$. Such a constraint can be obtained from Eq.~\eqref{eq:observable} for a generic observable, $O$, not commuting with $H$. We spell out the modified equations in Appendix~\ref{app:additional_constraints}. The optimization problem in Eq.~\eqref{eq:learning_cost} is then modified to
\begin{equation}\label{eq:additionalConstraints}
(\vec{c}^{\rm rec},\vec{d}^{\rm rec})=\argmin_{ (\vec{c},\vec{d}):\vec{d}\geq 0} \norm{\mqty( M (\vec{d}) \\ \xi M^\mathrm{\rm add})\vec{c} - \mqty(0 \\ \xi \vec{b}(\vec{d})) },
\end{equation}
where $M(\vec{d})=M_H+M_D(\vec{d})$ is as in Eq.~\eqref{eq:learning_cost}, and $M^{\rm add}$ and $\vec{b}(\vec{d})$ contain additional constraints. The \emph{auxiliary parameter} $\xi$ controls a penalty that is added for violating the condition determining $H$. Choosing this parameter large enough allows us to reconstruct a $\vec{c}^{\rm rec}\approx\vec{c}^H$, with the correct overall scale, as a unique solution of Eq.~\eqref{eq:additionalConstraints}. We emphasize, that by choosing the appropriate additional constraints, Eq.~\eqref{eq:additionalConstraints} can also be used to learn other conserved quantities of $H$.

\subsubsection{Learning the full Liouvillian}

Since we have dissipative dynamics, one expects that accurately learning the Hamiltonian also requires learning all the individual dissipation rates. This is, however, not necessarily the case. In some cases, different dissipative processes give the same contribution to Eq.~\eqref{eq:energy}, and therefore, cannot be distinguished by these conditions. More specifically, when inserting an ansatz into this condition, the resulting constraint matrices $\{M^{(k)}\}$ in Eq.~\eqref{eq:timeTraces}, where $k$ labels the different Lindblad operators, become linearly dependent. Then, the decomposition $M_D(\vec{d})=\sum_k d_k M^{(k)}$ is not unique. This dependence leads to symmetries in the cost function in Eq.~\eqref{eq:learning_cost}, such that the Hamiltonian $H$ can be learned exactly, for different dissipation rates $\boldsymbol{d}^\mathrm{rec}$.

In order to learn the full Liouvillian and resolve individual dissipation rates, one needs to add additional constraints in the same way as one does for excluding conserved quantities and for learning the overall scale. This can then be phrased as an optimization problem of the form of Eq.~\eqref{eq:additionalConstraints}, which we explain in detail in Appendix~\ref{app:additional_constraints}. This is a unique feature of Eq.~\eqref{eq:learningEq}. Below, we will compare Hamiltonian and Liouvillian learning based on Eq.~\eqref{eq:LearningEqEhrenfest}, which, generically, does not admit such symmetries, and Eq.~\eqref{eq:learningEq}, including additional constraints, where we will show how to choose these additional constraints to learn the full Liouvillian.

\section{Numerical Case Studies}\label{sec:examples}

In this section we want to illustrate the Hamiltonian and Liouvillian learning methods developed in the previous sections in the context of several experimentally relevant spin-Hamiltonians, more specifically the Ising model in Eq.~\eqref{eq:model1H}, and a long-range Hamiltonian with algebraically decaying interactions that is naturally found in trapped ion experiments~\cite{Monroe2021}. We will study the procedure of learning these Hamiltonians, in the presence of weak dissipation, using only a finite amount of numerically simulated measurement data. Moreover, we will aim to learn the Liouvillian, and in particular distinguish local from collective dissipation. As a preparation for the following discussion let us discuss some details of the learning procedure.

\subsection{Preliminaries}
To learn the Hamiltonian and Liouvillian of a many-body system, the constraint matrices in Eq.~\eqref{eq:LearningEqEhrenfest}, or Eq.~\eqref{eq:learningEq}, are estimated from a finite amount of experimental data. Therefore, we have to solve the corresponding optimization problem using their noisy estimates, e.g., $\tilde{K}_H=K_H+E$, where $E$ is an error matrix with random entries. Crucially, $E$ is experimentally inaccessible. Therefore, as an experimentally accessible figure of merit to asses the progress of our learning procedure, one may consider the minimum of the objective function in Eq.~\eqref{eq:learning_cost_Ehrenfest}, or Eq.~\eqref{eq:learning_cost}, given by the smallest singular value $\lambda_1$. In early stages of the learning, where shot-noise, represented by the matrix $E$, is the dominant source of error, one expects this minimum to decrease $\sim N_{\rm runs}^{-1/2}$ as we increase $N_{\rm runs}$, see Appendix~\ref{app:shot-noise} for details. In this regime, we will consider the reconstructed parameters $\vec{c}^{\rm rec}$ and $\vec{d}^{\rm rec}$, for which we compute error bars using bootstrapping methods. This allows us to identify dominant terms in the Hamiltonian and Liouvillian, thus learning the dominant operator structure. For a sufficient ansatz, this error will asymptotically converge to zero, while for an insufficient ansatz the error will be strictly bounded away from zero, i.e., it will~\emph{plateau} in later stages of the learning, where missing ansatz terms are the dominant source of error. Therefore, when observing a plateau we need to extend our ansatz by additional terms.

A fundamental tension in Hamiltonian and Liouvillian learning is the tradeoff between the number of parameters of an ansatz, and the number of measurements that are required to estimate these parameters from experimental data. While ansätze comprising many parameters allow us to learn many aspects of the dynamics, they require a large measurement budget to reduce shot noise. In contrast, in the case of a limited measurement budget, it is quite useful to reduce the number of parameters to just a few relevant ones. This is also physically justified as many experimentally relevant Hamiltonians can be effectively characterized by only a few parameters, which may not even increase with system size. Illustrative examples include translationally invariant systems, or systems exhibiting algebraically decaying interactions. While ansätze with few parameters only require a small measurement budget, they might be insufficient in the limit of an infinite number of measurements. However, as long as these insufficiencies are small, learning will be limited by shot noise. Therefore, our goal in the following is to find an ansatz comprising only few parameters and the dominant terms of the Hamiltonian and Liouvillian, which do not show any insufficiencies given our limited measurement budget.

To find this ansatz, we compare different ansätze by varying their operator content as well as their parametrization, i.e., the dependencies between parameters. We emphasize that \emph{reparametrization} can be done solely by classical post-processing of the data. More specifically, by transforming the vector of parameters, $\vec{c}$, via a \emph{parametrization matrix} $G$ to a new vector of parameters, $\vec{c}_G=G^T\vec{c}$, where $G$ encodes dependencies between parameters in $\vec{c}$. This leads to a transformation of the learning equations in Eqs.~\eqref{eq:LearningEqEhrenfest}, and~\eqref{eq:learningEq} [see Appendix~\ref{app:reparametrization} for technical details]. We will demonstrate in the following numerical simulations that the ratio of the smallest two singular values of the constraint matrices, $\lambda_1/\lambda_2$, serves as a quantifier for the learning error of an ansatz and a useful figure of merit to compare different ansätze, . This is motivated 
by the fact that the smallest singular value, $\lambda_1$, quantifies the learning error, e.g., the violation of generalized energy conservation in Eq.~\eqref{eq:energy}, and that a larger gap $\delta=\lambda_2$ tightens the upper bound on $\vert \sin(\theta) \vert$, where $\theta=\angle (\vec{c}^H,\vec{c}^{\rm rec})$ is the reconstruction error (see Appendix~\ref{app:shot-noise} for details). As we will see below, the gap is typically larger for ansätze with fewer parameters. Therefore, when comparing different ansätze, we choose the one which minimizes $\lambda_1/\lambda_2$. This results in an ansatz containing only few parameters and small missing terms, i.e., with an early shot-noise scaling, and a low plateau (see Fig.~\ref{fig:HL2} for an illustration).

As a final remark, let us note that in case the operators $O_i$, $h_j$ and $a_k$ are Pauli operators, the expectation values in Eqs.~\eqref{eq:constraintMatrix2} and~\eqref{eq:timeTraces} evaluate to $\alpha \expval{h_j}$, where $\alpha=0$ if $[l_k,h_j]=0$, and $\alpha=-4$ otherwise, and respectively for $O_i$. Therefore, in case the operators $\{O_i\}$, $\{h_j\}$ are few-body operators these expectation values as well as the commutator $[O_i,h_j]$ remain few-body Paulis. Thus, many of the operators commute and can thus be measured jointly.

Estimating the integrals in Eq.~\eqref{eq:constraint_vector_additional} requires measuring time-resolved expectation values of the corresponding operators over the duration of the quench, see Fig.~\ref{fig:HL2}. Here, we use the composite Simpson's rule~\footnote{It reads $\int_0^T \expval*{P}_t \d t \approx \frac{1}{3}\Delta t\left[\expval*{P}_{0} +4\sum_{k=1}^{N_t/2}\expval*{P}_{t_{2k-1}} +2\sum_{k=1}^{N_t/2-1}\expval*{P}_{t_{2k}}+ \expval*{P}_T\right]$.} which approximates an integral as a series of parabolic segments between $N_t$ equally spaced points $t_m=m\cdot \Delta t$, with $m\in 0,\dots,N_t$ and distance $\Delta t = T/N_t$.
In the following, we choose a sufficiently small time step $\Delta t$ to ensure that errors arising from discretizing the integral can be disregarded in comparison to shot-noise errors~\footnote{Indeed, for fixed time $T$ one obtains $\int_0^T\expval*{h}_t\d t=I(\Delta t)+K\Delta t^4$, where $I(\Delta t)$ is the integral approximation, and $K$ is a constant. When expressed in terms of $N_t$ this reads $I(N_t)+\tilde{K}/N_t^4$. However, as $I(N_t)$ can only be estimated from data, one obtains $I(N_t)=\tilde{I}(N_t)+\epsilon$, only up to a statistical error $\epsilon$. As $I(N_t)$ is linear, the variance of $\epsilon$ roughly scales like $1/(N_tN_s)$, where $N_s$ is the number of shots per time-point.}.

\begin{figure*}
\centering
\subfloat[\label{fig:example1a}]{
  \includegraphics[width=0.25\linewidth]{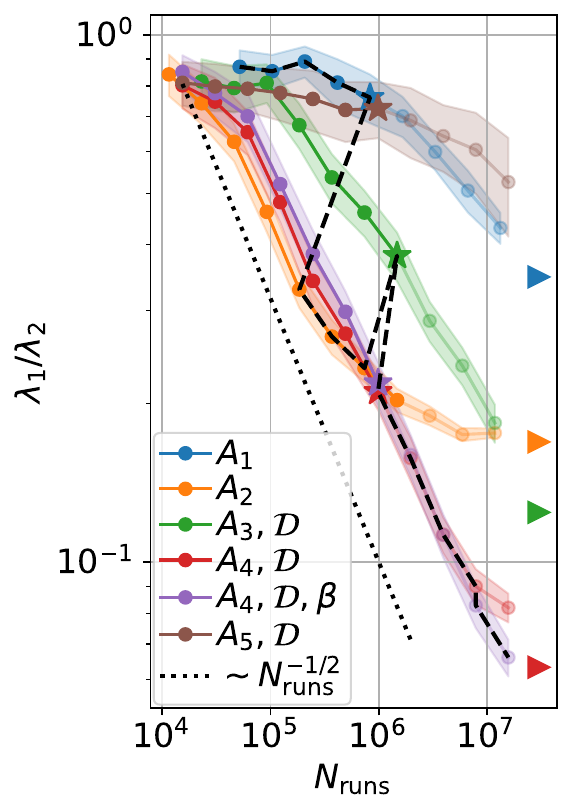}
}\hspace*{1cm}
\subfloat[\label{fig:example1b}]{
  \includegraphics[width=0.57\linewidth]{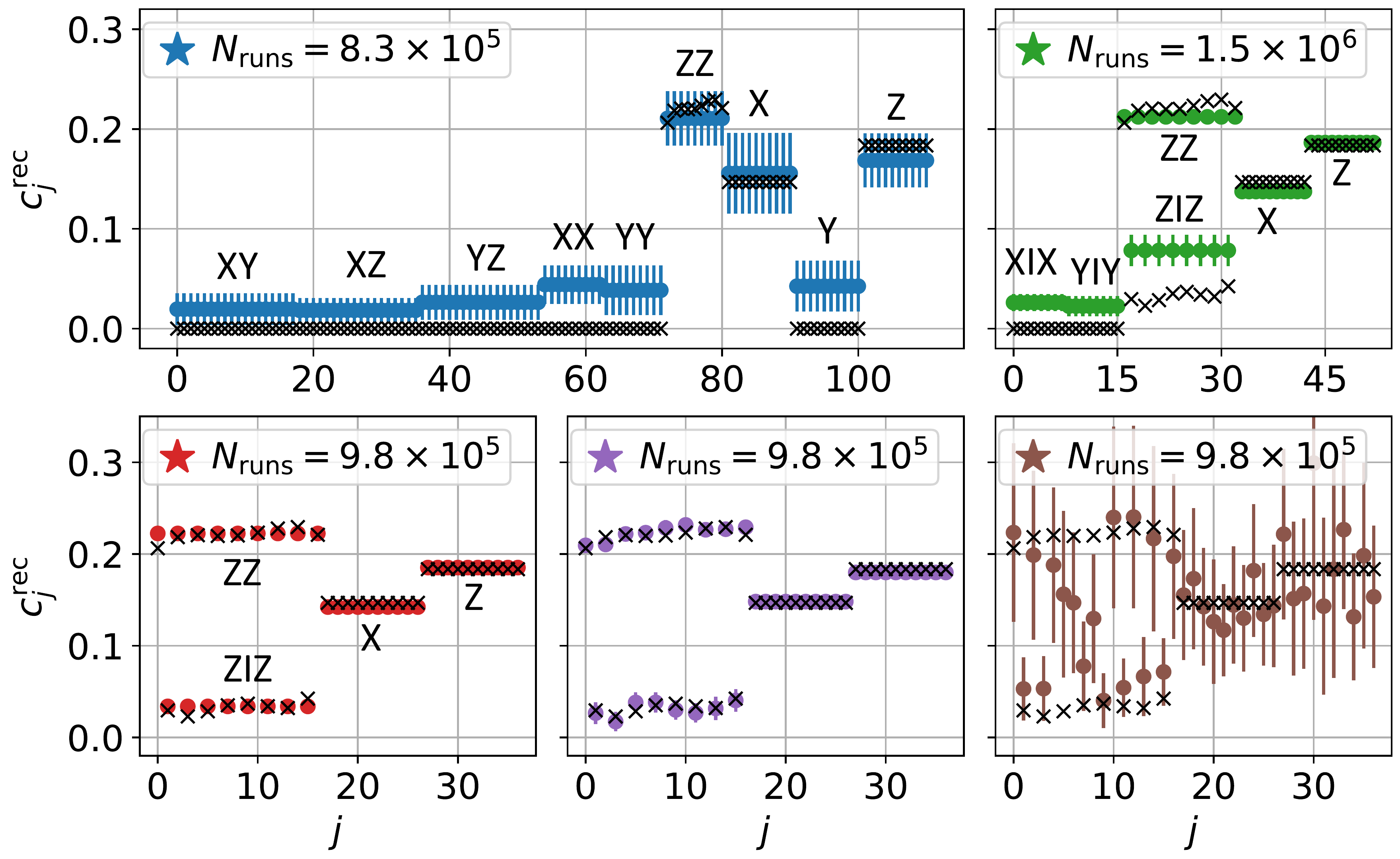}
}
\caption{Hamiltonian learning in presence of weak dissipation. (a) The learning error $\lambda_1/\lambda_2$ as a function of the number of runs, $N_{\rm runs}$, of the simulator for different ansätze described in the main text. The colored triangles indicate the asymptotic values for $N_{\rm runs}\to\infty$. The dashed line follows the learning procedure described in the main text.
(b) Snapshots of the (normalized) reconstructed Hamiltonian parameters, $c^{\rm rec}_j$, for some of the ansätze described in the main text. The capital letters denote groups of operators in the ansatz, and within each group the site indices of the coefficients are sorted increasingly from left to right, e.g., the coefficients in group $XY$ correspond to the operators $\sigma^x_{k} \sigma^y_{k+1}+\sigma^y_{k} \sigma^x_{k+1}$ in Eq.~\eqref{eq:A1}, and similarly for all other groups of operators. Black crosses indicate the true parameters of the model Hamiltonian as defined below. The point at which these snapshots are taken in terms of $N_\mathrm{runs}$ is also indicated by asterisks in the left panel. All error bars are computed via bootstrapping as explained in Appendix~\ref{app:errorbars}, using $80$ samples.  
{\em Model parameters:} The Hamiltonian parameters in Eq.~(\ref{eq:model1}) are chosen as $B_x = 4/5 B_z$. Moreover, the coupling strengths are described by $4$-th order polynomials of the form $J_{i,i+1}^z = B_z \sum_{l=1}^5 a_l \cdot x_{i,i+1}^{l-1}$, and $J_{i,i+2}^z = B_z \sum_{l=1}^5 b_l \cdot x_{i,i+2}^{l-1}$ where $x_{i,j}=[(i+j)-(N+1)]/N$. We choose the following coefficients: $\boldsymbol{a}=(6/5,1/20,1/5,0,-2/5)$ and $\boldsymbol{b}=(1/5,1/20,-2/5,0,4/5)$. The dissipation rates are chosen such that $(\gamma_+,\gamma_-,\gamma_z) = (1,1.5,2) \times 10^{-2} B^z \ll \Vert H \Vert$. To estimate the integrals in Eq.~\eqref{eq:timeTraces}, we use $B_z \Delta_t = 1/64$, i.e., $64$ equidistant times, where at each time we spend only $1/64$ of the measurements per basis, compared to the quench times $B_z T \in [0,0.5,1]$. Expectation values at $t=0$ are calculated exactly from the initial state.}
\label{fig:example1}
\end{figure*}

\subsection{Learning from generalized energy conservation} \label{sec:example1}

To begin, we illustrate Hamiltonian and Liouvillian learning from generalized energy conservation based on Eq.~\eqref{eq:learningEq}. We assume that the analog quantum simulator is governed by a master equation of the following form
\begin{equation} \label{eq:model1} 
    \frac{\d}{\d t}\varrho = -i [H,\varrho] + \sum_{\eta\in\{+,-,z\}}\frac{\gamma_\eta}{2} \sum_{k=1}^{N} \qty( [\sigma_k^\eta \varrho, \sigma_k^{\eta,\dagger}] + \mathrm{h.c.}).
\end{equation}
Here, $H$ is the Ising Hamiltonian of Eq.~\eqref{eq:model1H} and the Lindblad operators $\sigma_+$, $\sigma_-$, and $\sigma_z$ represent spontaneous absorption, spontaneous emission, and local dephasing, respectively. In this model case, the Hamiltonian couplings and dissipation rates are chosen such that the dominant terms are the nearest-neighbor couplings and fields of the Ising Hamiltonian, while the sub-dominant terms are next-nearest-neighbor couplings in the Hamiltonian as well as the dissipative processes. Moreover, we included small spatial variations in the couplings of the Hamiltonian. We summarize the choice of parameters in the caption of Fig.~\ref{fig:example1}.

In the following, our goal will be to learn the above-described model Hamiltonian and Liouvillian from simulated quench experiments. We will limit the total number of simulated runs $N_\mathrm{runs} \approx 10^6$, which is a reasonable limit for experiments with trapped ions. We will illustrate how $\vec{c}^{\rm rec}$, including its error bars that we obtain via bootstrapping, and the learning error $\lambda_1/\lambda_2$ can be used in this scenario to identify the operator content and relevant parameters of the model Hamiltonian and Liouvillian.

\paragraph*{1. Identifying the dominant terms of the Hamiltonian.}

In a first step, we seek to identify the dominant terms in the Hamiltonian $H$, assuming (prior knowledge) that dissipation is typically weak compared to the Hamiltonian. Therefore, in this first step we will only include an ansatz for the Hamiltonian and no ansatz for the Liouvillian. As, interactions are typically finite range, we generically expect the dominant terms to be nearest-neighbor terms. Moreover, in quantum simulation of condensed matter models one expects more or less homogeneous couplings (translational invariance). Therefore, we start with the following ansatz for the Hamiltonian
\begin{multline}\label{eq:A1}
    A_1 = c_{xx}\sum_{k=1}^{N-1} \sigma_k^x\sigma_{k+1}^x+ c_{xy}\sum_{k=1}^{N-1} (\sigma_k^x\sigma_{k+1}^y+ \sigma_k^y\sigma_{k+1}^x) \\+ \dots  + c_{x}\sum_{k=1}^{N} \sigma_k^x + c_{y}\sum_{k=1}^{N} \sigma_k^y + c_{z}\sum_{k=1}^{N} \sigma_k^z ,
\end{multline}
comprising all nearest-neighbor interactions, i.e., $\qty{xx,yy,zz,xy,xz,yz}$, with spatially homogeneous coefficients. As commuting operators can be measured jointly, the operators in $A_1$ can be measured by classically post-processing data obtained from measuring the following nine product operators, independently of the system size: $\sigma_x^{\otimes N}$, $\sigma_y^{\otimes N}$, $\sigma_z^{\otimes N}$, and six operators of the form $\sigma_a\otimes \sigma_b\otimes \sigma_a\otimes \sigma_b\otimes \cdots$, for all combinations $a \neq b$ of distinct Pauli operators.

In Fig.~\ref{fig:example1a} we plot $\lambda_1/\lambda_2$ for the ansatz $A_1$ as function of $N_\mathrm{runs}$ (blue line). At small $N_\mathrm{runs}$ one expects to observe a decrease of $\lambda_1/\lambda_2$ with $N_\mathrm{runs}$ indicating the errors are dominated by shot noise. As the ansatz $A_1$ is insufficient, i.e., misses terms present in Eq.~\eqref{eq:model1H}, one expects a plateau in $\lambda_1/\lambda_2$ for large $N_{\rm runs}$. This plateau appears beyond our maximum measurement budget at $N_{\rm runs}\approx10^7$ experimental runs, which is indicated by a blue triangle on the y-axis in Fig.~\ref{fig:example1a}. 
Nevertheless, at $N_\mathrm{runs} \approx 10^6$ the reconstructed parameters $\vec{c}^{\rm rec}$ in Fig.~\ref{fig:example1b} (blue data) identify the dominant terms in the Hamiltonian, i.e., nearest-neighbor $\sigma_z\otimes\sigma_z$ interactions and fields in $\sigma_x$ and $\sigma_z$ direction.

At this point, we cannot rule out the presence of the other terms in the operator ansatz $A_1$. However, as the values of the non-dominant terms are much smaller than the ones of the dominant terms (see Fig.~\ref{fig:example1b}, blue data), we remove all non-dominant terms from the ansatz in the next step. It should be noted that, in case those terms were present in the Hamiltonian, we would encounter a plateau in $\lambda_1/\lambda_2$ at a later stage of the learning process. We will later see that this is, however, not the case (as dissipative terms are still missing in our ansatz).
On the other hand, if the dominant terms identified from $\boldsymbol{c}^\mathrm{rec}$ of $A_1$ constitute a good approximation to the Hamiltonian, we expect $\lambda_1/\lambda_2$ to decrease. This is indeed the case as the reparametrized ansatz,
\begin{equation}
\begin{split} \label{eq:A2}
    A_2 = c_{zz}\sum_{k=1}^{N-1} \sigma_k^z\sigma_{k+1}^z+ c_{x}\sum_{k=1}^{N} \sigma_k^x + c_{z}\sum_{k=1}^{N} \sigma_k^z
\end{split}
\end{equation}
leads to a much smaller value of $\lambda_1/\lambda_2$. We emphasize that learning with the ansatz $A_2$ only requires measurements of $\sigma_x^{\otimes N}$ and $\sigma_z^{\otimes N}$. Therefore, some of the measurements performed for $A_1$ can be recycled for learning the parameters of $A_2$. 
By further measuring, a plateau in $\lambda_1/\lambda_2$ starts to emerge at $N_\mathrm{runs}\approx 3\times 10^5$, as shown in Fig.~\ref{fig:example1a} (orange line).
At this point we have identified all the dominant terms in $H$.
To continue the learning process we need to extend the operator ansatz. Therefore, in the next step we want to learn subdominant terms of $H$, as well as the Liouvillian.

\paragraph*{2. Learning the sub-dominant terms of the Hamiltonian, and learning the Liouvillian.}

One expects, a priori, that when learning smaller terms of the Hamiltonian, weak dissipative effects become relevant. Therefore, we now include an ansatz for the Liouvillian
\begin{equation}
    \mathcal{D}(\varrho)=\sum_{\eta\in\{+,-,z\}}\frac{d_\eta}{2} \sum_{k=1}^{N} \qty( [\sigma_k^\eta \varrho, \sigma_k^{\eta,\dagger}] + \mathrm{h.c.}),
\end{equation}
and extend our ansatz for the Hamiltonian by next-nearest neighbor couplings. As before, we choose our ansatz to be spatially homogeneous, which leads to
\begin{multline}
    A_3 = A_2 + c_{x1x}\sum_{k=1}^{N-2} \sigma_k^x\sigma_{k+2}^x \\ +c_{y1y}\sum_{k=1}^{N-2} \sigma_k^y\sigma_{k+2}^y+c_{z1z}\sum_{k=1}^{N-2} \sigma_k^z\sigma_{k+2}^z.
\end{multline}
This ansatz for the Hamiltonian and Liouvillian requires measurements of the form $\sigma_x^{\otimes N}$, $\sigma_y^{\otimes N}$ and $\sigma_z^{\otimes N}$, which means that all of the data taken for $A_2$ can be reused for learning the parameters of $A_3$.
Note that including dissipation requires the same measurement bases, but at various times to estimate the integrals in Eq.~\eqref{eq:timeTraces}. 

An ansatz for the dissipation and next-nearest-neighbor terms now leads to a lower plateau of $\lambda_1/\lambda_2$ compared to $A_2$. However, this plateau appears for $N_{\rm runs}\geq 10^7$, see Fig.~\ref{fig:example1a} (green line), which exceeds our assumed measurement budget. Nevertheless, the reconstructed parameters of the Hamiltonian, $\vec{c}^{\rm rec}$, at $N_{\rm runs}\approx 10^6$, suggest that the next-nearest-neighbor $\sigma_z\otimes\sigma_z$ couplings are larger compared to other next-nearest-neighbor terms, see Fig.~\ref{fig:example1b} (green data). Therefore, we reparametrize our ansatz for the Hamiltonian to
\begin{equation}
    A_4 = A_2 + c_{z1z}\sum_{k=1}^{N-2} \sigma_k^z\sigma_{k+2}^z.
\end{equation}
Indeed, this leads to a significantly smaller ratio $\lambda_1/\lambda_2$ compared to $A_2$ and $A_3$, see Fig.~\ref{fig:example1a} (red line). Moreover, the small error bars in the reconstructed parameters in Fig.~\ref{fig:example1b} (red data) show that $N_\mathrm{runs} = 10^6$ is already sufficient for learning $A_4$.
Note that at this point, we have identified all dominant and subdominant terms. However, so far we have not learned a possible spatial structure of the couplings in $H$.

\paragraph*{3. Learning the spatial variations of the couplings.}

As can be seen in Fig.~\ref{fig:example1a} (red line), the plateau in $A_4$ is reached at $N_\mathrm{runs} \approx 10^7$, which again exceeds our available measurement budget. 
Therefore, we are not allowed to conclude that ansatz $A_4$ is insufficient to describe the Hamiltonian in Eq.~\eqref{eq:model1H} by only considering the ratio $\lambda_1/\lambda_2$. However, at this point, one can choose to test other ansätze and compare their corresponding reconstructed parameters, or to see if smaller values of $\lambda_1/\lambda_2$ are achieved. As one typically expects small spatial variations in the couplings of $H$, and $A_4$ was already a good approximation of $H$, we want to test for spatial variations on top of the spatially homogeneous ansatz $A_4$. To this end, we simply treat the parametrization of the ansatz $A_4$ as a regularization of the cost function, see Appendix~\ref{app:reparametrization} for technical details. That is, we start with the optimization problem in Eq.~\eqref{eq:learning_cost}, and add a penalty term with control parameter $\beta$ for deviating from a given parametrization. Choosing $\beta \gg 1$ imposes the exact parametrization of $A_4$, and we then successively decrease $\beta$ until one observes the emergence of spatial variations, and the corresponding error bars, see Fig.~\ref{fig:example1b} (purple data). This ensures that our learning process is dominantly limited by shot noise, and not by having too few parameters in the ansatz.

With a larger measurement budget one could then successively increase $N_\mathrm{runs}$ while decreasing $\beta$, until the desired accuracy is reached, which is reminiscent of Bayesian learning~\cite{Evans2019,Mehta2019}. Therefore, we conclude that we have indeed successfully determined the operator content and parameters of the Hamiltonian and Liouvillian in our simulated experiment. 

As a final step, we may compare the reconstructed parameters of our learning procedure to the ones obtained from a "naive" ansatz of the form
\begin{multline}
A_5 = \sum_{k=1}^{N-1} c_{zz}^{(k)} \sigma_k^z\sigma_{k+1}^z+ \sum_{k=1}^{N-2} c_{z1z}^{(k)} \sigma_k^z\sigma_{k+2}^z\nonumber \\  \qquad + \sum_{k=1}^{N} c_{x}^{(k)} \sigma_k^x +\sum_{k=1}^{N}  c_{z}^{(k)} \sigma_k^z,
\end{multline}
that has a total of $4N-3$ parameters. One notices that our strategy leads to a much more accurate reconstruction, cf. Fig.~\ref{fig:example1b} (brown data).

\begin{figure*}
\centering
    \includegraphics[width=\linewidth]{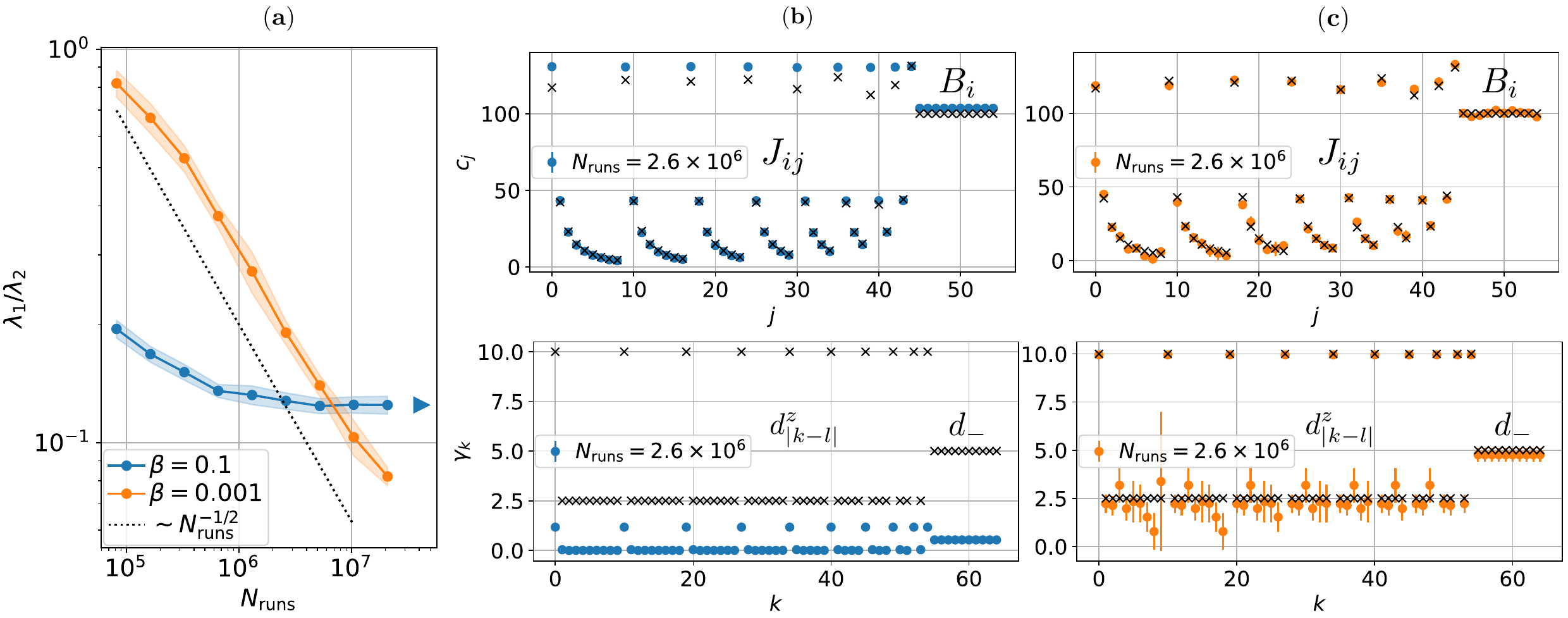}
    \caption{Learning collective dissipation using random measurements. (a) The ratio $\lambda_1/\lambda_2$ of the matrix $(M,-b)$ for different regularization parameters $\beta$. (b,c) Snapshots of the learned Hamiltonian parameters, $c^{\rm rec}_j$, corresponding to the parameters $J_{kl}$ and $B_i$ defined in Eq.~\eqref{eq:model_hamiltonian_FlipFlop}. The required measurement budget is also indicated in the left panel. Black crosses correspond to the parameters of the model Hamiltonian specified below. All error bars are computed via bootstrapping as explained in Appendix~\ref{app:errorbars}, using $40$ samples. The time-traces in Eq.~\eqref{eq:timeTraces} are evaluated using $B_z \Delta_T = 1/20$, i.e., using $20$ time-steps, using the same amount of random measurements at each time up to a time of $B_z T = 0.5$. Expectation values at $t=0$ are calculated exactly from the initial state.
    {\em Model parameters:}
For the Hamiltonian we choose parameters as specified in Eq.~\eqref{eq:parameters_model2_ideal} with $B_z=100$, $J_0= 6/5 B_z$ and $\alpha=1.5$. Moreover, in the Liouvillian, we choose $\gamma_- = B_z/20 $ in Eq.~\eqref{eq:decay}, as well as $\Gamma_{kl}^z = \gamma_k^z \delta_{kl} +  \Gamma_0$
with $\Gamma_0=B_z/40$ and $\gamma_k^z = 3 \Gamma_0$ in Eq.~\eqref{eq:collective_dissipation}.}
    \label{fig:example3_dissipation}
\end{figure*}

\subsection{Learning from generalized Ehrenfest theorem}\label{sec:ex3_collective}

In this section, we will learn the Hamiltonian and Liouvillian of a model system from the generalized Ehrenfest theorem in Eq.~\eqref{eq:LearningEqEhrenfest}. As this method requires the estimation of expectation values of many observables, $\{O_i\}$, we will use the randomized measurement toolbox~\cite{Elben2022} to obtain the necessary data. This will also allow us to demonstrate Hamiltonian learning in large systems of tens of qubits, and provide evidence that given a fixed number of shots the reconstruction error is independent of the system size, if the number of Hamiltonian parameters can be kept constant.

We consider a model system involving long-range spin-spin interactions, as realized in trapped-ion setups. The effective Hamiltonian is given by~\cite{Monroe2021}
\begin{equation} \label{eq:model_hamiltonian_FlipFlop}
\begin{split}
    H_\mathrm{XY} = \sum_{i<j}^N J_{ij} \left( \sigma_i^x \sigma_j^x + \sigma_i^y \sigma_j^y \right) + \sum_{i=1}^N B_i \sigma_i^z ,
\end{split}
\end{equation}
with interaction strengths $J_{ij}$ and site-dependent magnetic field $B_i$. In an idealized description, it is often assumed that
\begin{equation}
\label{eq:parameters_model2_ideal}
\begin{split}
    J_{ij} = \frac{J_0}{\vert p(i) - p(j) \vert^\alpha}, \quad
    B_i = B_z,
\end{split}
\end{equation}
where $p(i)$ is the position of the $i$-th ion and with tunable $0 \leq \alpha \leq 3$. To add features to the model, we assume that the ions are randomly shifted from their equilibrium positions, i.e., $p(i) = i + r_i$, where each $r_i$ is sampled uniformly from $[-0.05,0.05]$. Models with random ion positions are studied, for example, in the context of topological defects in the Frenkel-Kontorova model~\cite{Chelpanova2024,Garcia-Mata2007}. 

To account for spontaneous emission of trapped ions, we add the spatially homogeneous Lindblad terms
\begin{equation} \label{eq:decay}
    \mathcal{L}^{-}(\varrho)=\frac{\gamma^-}{2}\sum_{k=1}^{N} \left( [\sigma_k^- \varrho, \sigma_k^{-,\dagger}] + [\sigma_k^-, \varrho \sigma_k^{-,\dagger}] \right).
\end{equation}
Moreover, we consider the presence of dephasing terms originating from the presence of a fluctuating, classical magnetic field, that leads to shifts of the energy levels of the $k$-th ion proportional to $B_k^\mathrm{cl}(t)$. It can be shown that in the white-noise limit, i.e., with correlation function $\langle B_k^\mathrm{cl}(t) B_l^\mathrm{cl}(t^\prime) \rangle = \Gamma_{kl}^z \delta(t-t^\prime)$, this leads to Lindblad terms of the form
\begin{equation} \label{eq:collective_dissipation}
\begin{split}
 \mathcal{L}^z(\varrho) = \sum_{k,l=1}^{N} \Gamma_{kl}^z \Big\{ \sigma_k^z \varrho \sigma_{l}^z - \frac{1}{2} (\sigma_k^z \sigma_{j}^z \varrho + \varrho \sigma_k^z \sigma_{j}^z ) \Big\}.
\end{split}
\end{equation}
In particular, uncorrelated magnetic field fluctuations lead to a diagonal matrix $\Gamma_{kl}^z = \gamma_{k}^z \delta_{kl}$, whereas global fluctuations, with $B_k^\mathrm{cl}(t)=B^\mathrm{cl}(t)$ for all $k$, lead to a constant matrix $\Gamma_{kl}^z = \Gamma_0$. 
In the present example we will, for simplicity, only consider uncorrelated and global fluctuations, which lead to a matrix
\begin{equation}
\label{eq:dissipation_matrix_collective}
    \Gamma_{kl}^z = \gamma_k^z \delta_{kl} +  \Gamma_0,
\end{equation}
[see also Fig.~\ref{fig:example3_dissipation} for the specific choices of parameters]. We note, that in experiments with, for instance, long ion strings, or local magnetic fields one expects a more complicated structure of the matrix $\Gamma_{kl}^z$.

\paragraph*{1. Hamiltonian and Liouvillian learning using randomized measurements.}

In Section~\ref{sec:example1}, we already illustrated learning of the operator content, and the relevant parameters of the Hamiltonian from minimal assumptions. Here, we want to focus on distinguishing local from collective dephasing, and analyse the scaling of the reconstruction error.
Therefore, we start with a parametrized ansatz, that already incorporates the algebraically decaying spin-spin interactions in $H_\mathrm{XY}$, that one would expect to find in a trapped-ion experiment, as well as a constant magnetic field. We choose as an ansatz for the Hamiltonian
\begin{equation}\label{eq:ansatzAXY}
    A(\vec{c}) = \sum_{i<j} \frac{c_0}{\vert i-j\vert^{\alpha}} (\sigma_i^x \sigma_{j}^x + \sigma_i^y \sigma_{j}^y) +  b \sum_k \sigma_k^z,
\end{equation}
which depends non-linearly on the parameter $\alpha$, and hence, will require non-linear minimization of $\lambda_1$ over $\alpha$ as explained in Appendix~\ref{app:reparametrization}.
As an ansatz for the Liouvillian we choose
\begin{equation}
\label{eq:ansatz_dissipators3}
    \mathcal{D}(\varrho) =\frac{d_-}{2} \sum_{k=1}^{N} \qty( [\sigma_k^- \varrho, \sigma_k^{-,\dagger}] + \mathrm{h.c.}) + \sum_{k,l=1}^N d^z_{\vert k-l\vert}  \mathcal{D}^z_{kl}(\varrho),
\end{equation}
that includes single-spin spontaneous decay with jump operator $\sigma_k^-$, and single-spin as well as multi-spin collective dephasing $\mathcal{D}^z_{kl}$ with operator structure as described in Eq.~\eqref{eq:collective_dissipation}.

To obtain constraint operators, $\{O_i\}$, we proceed as follows; we choose a set of random Pauli measurements, $\{P_{\vec{\alpha}}\}=\{\sigma^{\alpha_1}\otimes \dots \otimes\sigma^{\alpha_N}\}$, with $\alpha_i \in \{x,y,z\}$, which we measure at consecutive times from $t=0$ up to $t=T$, similar as before (see Fig.~\ref{fig:example3_dissipation} for details). From this set of Pauli measurements we extract all one- and two-qubit operators that are contained in $\{P_{\vec{\alpha}}\}$, as well as their commutators with the ansatz operators and jump operators (in case those are contained). For these operators and commutators we extract expectation values to obtain the data necessary to estimate the constraint matrix in Eq.~\eqref{eq:LearningEqEhrenfest}. Choosing the set $\{P_{\vec{\alpha}}\}$ large enough ensures that the necessary operators are contained in this set.

The results are presented in Fig.~\ref{fig:example3_dissipation}.
In Fig.~\ref{fig:example3_dissipation}(a) we show the ratio of the two smallest singular values of the matrix $(K_H,K_D,-\boldsymbol{b})$, which can be obtained by rewriting Eq.~\eqref{eq:LearningEqEhrenfest}, where $K_H, K_D$, and $\boldsymbol{b}$ are the constraint matrices and vector respectively. The quantity $\lambda_1/\lambda_2$ shows a similar scaling behavior in early and late stages of the learning procedure as the analog quantity in Section~\ref{sec:example1}. Moreover, it correctly recognizes the insufficiency of the translation invariant ansatz ($\beta=0.1$, blue), as well as the larger number of parameters of the almost unparametrized ansatz ($\beta=0.001$, orange). In Fig.~\ref{fig:example3_dissipation}(b) we show the learned Hamiltonian parameters (top) and dissipation rates (bottom) for both ansätze, and a total measurement budget of $N_\mathrm{runs}=2.6\times10^6$, in comparison to the true values (black). For the translation invariant ansatz (blue) one notices that while the reconstructed parameters of the Hamiltonian roughly resemble the true parameters, the dissipation rates turn out wrong. We have numerically verified that this artefact does not disappear for larger measurement budget. However, by lowering the penalty $\beta$ for deviating from translation inference the solution $\vec{c}^{\rm rec}$ correctly accounts for disorder, and the dissipation rates are learned accurately.

\paragraph*{2. Comparison between learning form Ehrenfest constraints and energy conservation. }
To make a fair comparison between Hamiltonian and Liouvillian learning from energy conservation and the Ehrenfest theorem, we have learned the same model system using Eq.~\eqref{eq:learningEq}. This, however, also requires the measurement of additional constraints to learn the overall scale, and to distinguish local from collective dephasing due to symmetries of $H_{XY}$. The results are discussed in Appendix~\ref{app:HXYfromEnergyConservation}. One observes, that in a system with conserved quantity, learning from the Ehrenfest theorem in Eq.~\eqref{eq:LearningEqEhrenfest} requires less measurements, when compared to learning from generalized energy conservation in Eq.~\eqref{eq:learningEq}. For instance, verifying the presence of collective dephasing using the Ehrenfest theorem requires $N_{\rm runs}\approx 10^6$ (cf. Figure~\ref{fig:example3_dissipation}), compared to $N_{\rm runs}\approx 10^7$ when using energy conservation (cf. Figure~\ref{fig:example2_diss} in Appendix~\ref{app:HXYfromEnergyConservation}).

\begin{figure*}[ht]
    \centering
    \includegraphics[width=\linewidth]{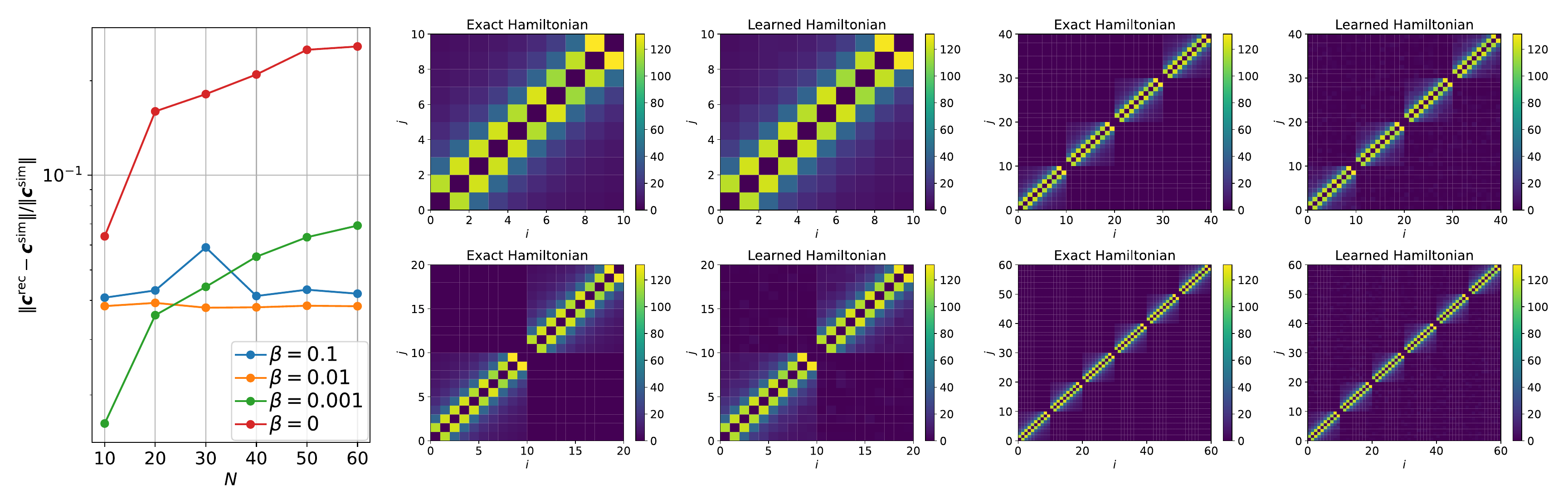}
    \caption{Hamiltonian learning with random measurements using a fixed total number of random measurements $N_\mathrm{runs}=1.9 \times 10^6$. The integrals in Eq.~\eqref{eq:timeTraces} are solved using $\Delta_T= T/10$, i.e., using $10$ equidistant time-steps, where at each time we use the same amount of measurements of $N_{\rm bases}=2000$ random bases, with a total time of $B_z T = 0.1$ for $B_z$ as in Fig.~\ref{fig:example2}. Expectation values at $t=0$ are calculated exactly from the initial state. (a) The error in the learned parameters as function of system size for different regularization strengths $\beta$. (b) Learned parameters for different system sizes $N \in [10,20,40,60]$ for a fixed regularization strength $\beta=0.001$.
    {\em Model parameters:} We choose the same interaction strengths $J_{ij}$ as in Fig.~\ref{fig:example2}, but setting the dissipation to zero and $B_z=0$.
    }
    \label{fig:HL_with_shadows}
\end{figure*}

\paragraph*{3. Scaling to larger system sizes.} \label{sec:ex3_large_scale}

We now want to study the scaling behavior of the reconstruction error of the Hamiltonian parameters, by varying system size from $N=10$ up to $N=60$ spins, for a Hamiltonian that is dominantly described by a constant number of parameters. In order to simplify the presentation and to reduce the classical computation time needed for simulating the measurements we make the following modifications to our model system: (i) We set the magnetic field and the dissipation rates to zero. (ii) We consider a Hamiltonian of the form
\begin{equation} \label{eq:H_subsystems}
    H = \sum_{k=1}^{N_\mathrm{sub}} \sum_{i\neq j = 1}^{N} J_{ij} \left( \sigma_{(kN+i)}^x \sigma_{(kN+j)}^x + \sigma_{(kN+i)}^y \sigma_{(kN+j)}^y \right),
\end{equation}
consisting of $N_{\rm sub}$ non-interacting subsystems of $N=10$ spins each, which makes it easy to simulate on a classical computer.

To learn the Hamiltonian in Eq.~\eqref{eq:H_subsystems} for the different system sizes, we want to compare leaning without parametrization, where the number of parameters in the ansatz grows with systems size, to learning using a parametrized ansatz where the number of parameters is kept constant with system size. Therefore, we choose an ansatz
\begin{multline} \label{eq:ansatz_subsystems}
    A(\boldsymbol{c}) = \sum_{k=1}^{N_\mathrm{sub}} \sum_{i\neq j = 1}^{10} \frac{c_{0}}{\vert i - j \vert^\alpha} \left( \sigma_{i+k}^x \sigma_{j+k}^x + \sigma_{i+k}^y \sigma_{j+k}^y \right) \\
    + \sum_{k \neq l =1}^{N_\mathrm{sub}} \sum_{i\neq j=1}^{10} c_1 \left( \sigma_{i+k}^x \sigma_{j+l}^x + \sigma_{i+k}^y \sigma_{j+l}^y \right),
\end{multline}
with a constant number of parameters, and that depends non-linearly on the parameter $\alpha$ similar to Eq.~\eqref{eq:ansatzAXY}. 
As in the previous example the parametrization of this ansatz is not strictly enforced, but depends on the chosen regularization strength $\beta$ (see Appendix~\ref{app:reparametrization} for technical details).
We emphasize that for $\beta=0$ we are learning from an unparametrized ansatz, with an independent parameter for each operator in the ansatz. In this case, our reconstructed Hamiltonian does not have the subsystem-structure of the true Hamiltonian in Eq.~\eqref{eq:H_subsystems} inherently built-in; rather, this structure is learned by including all possible inter-subsystem interaction terms.

We learn the Hamiltonian in Eq.~\eqref{eq:H_subsystems} for $N=10,\dots,60$ systems. We choose a total of $2000$ random Pauli measurements for each of a total of $11$ time-steps up to $T=1$, leading to a total measurement budget of $N_{\rm runs}=1.5 \times 10^6$. 
For the constraint matrix $M$ and vector $b$ in Eq.~\eqref{eq:learning_cost_Ehrenfest} we choose the constraints to be all possible one- and two-qubit constraint operators, and evaluate the corresponding constraint matrix $M$ and vector $b$. We then determine the learned coefficients $\boldsymbol{c}^\mathrm{rec}$ from Eq.~\eqref{eq:learning_cost_Ehrenfest}.

The result is shown in Fig.~\ref{fig:HL_with_shadows}. In Fig.~\ref{fig:HL_with_shadows}(a) we show the relative error of the learned parameters defined by
\begin{equation}
    \Delta \boldsymbol{c} = \Vert \boldsymbol{c}^\mathrm{rec} - \boldsymbol{c}^\mathrm{sim} \Vert / \Vert \boldsymbol{c}^\mathrm{sim} \Vert
\end{equation}
as a function of the system size $N$. We find that for weakly, or unparamatrized ansätze, for which the number of learned parameters grows with system size, the reconstruction error grows only moderately. This is likely explained by the fact that in larger systems a single round of measurements extracts more information as more correlation functions can be measured in a single shot.  Moreover, for parametrized ansätze with a system-size independent number of parameters the reconstruction error also remains independent of the system-size.

\section{Conclusion and Outlook}

In this study, we have devised and compared different methods for learning the Hamiltonian and Liouvillian in the analog quantum simulation of many-body systems. Our work applies to a scenario where one has direct access to the quantum device, however, in the literature, other scenarios have been considered, where one does not have direct access to the quantum device, see, e.g., Refs.~\cite{Hangleiter2017,Liu2024}. Our method is applicable in a regime of experimental relevance where the dissipation is weak compared to the coherent evolution. Our protocols are based on quench experiments, where initial product states evolve under coherent and dissipative dynamics, and the resulting state is measured in a product basis. Hamiltonian and Liouvillian learning can be understood as a sample efficient process tomography of quantum simulators. The learning methods begin with an ansatz for the operator structure of the Hamiltonian and Liouvillian. The quality of this ansatz can be monitored by measuring the learning error. Our strategy encompasses the reparametrization of the ansatz. This typically allows data recycling from previous measurement runs, but requires additional classical post-processing. This approach enables us to identify step by step first the dominant operator content of the Hamiltonian and Liouvillian, and successively sub-dominant terms within a limited measurement budget. A distinctive feature of learning from generalized energy conservation is that we can learn the Hamiltonian without the necessity to learn the entire Liouvillian, thus reducing the number of parameters to be learned. However, we demonstrated that additional constraints can be employed to learn the entire Liouvillian and to ascertain the overall scale of the Hamiltonian. Furthermore, these additional constraints can facilitate the learning of the Hamiltonian even when there are conserved quantities as operators commuting with the system Hamiltonian. We demonstrate how randomized measurements~\cite{Elben2022} can be utilized to learn the Hamiltonian and Liouvillian from the generalized Ehrenfest theorem, as this requires the estimation of many different correlation functions. Moreover, this allowed us to numerically demonstrate Hamiltonian learning for large systems consisting of tens of spins. While the focus of the present paper has been on spin models, the central ideas of Hamiltonian and Liouvillian learning also carry over to Bose and Fermi Hubbard models.

Extensions of the present work should consider scenarios where the experimental Hamiltonian (and Liouvillian) involves a large number of small terms, which, e.g., emerge as corrections in effective many-body spin models in a low-energy description. Such a formulation might involve a statistical description as learning of an ensemble of Hamiltonians. Along similar lines, Hamiltonian learning might also account for slow drifts of experimental Hamiltonians and Liouvillians. Considering alternative viewpoints, Hamiltonian and Liouvillian learning can also be phrased in the language of Bayesian inference, similar to Ref.~\cite{Evans2019}, establishing an interesting link between techniques of parameter estimation in multi-parameter quantum metrology, and optimal sensing with finite measurement budgets.
Finally, exploring alternative routes should include using (short-range) entangled states as inputs, which may be easily prepared in experiments.

\begin{acknowledgments}

We would like to thank Manoj K. Joshi, and Christian Roos for discussions and valuable feedback on the manuscript. TK would like to thank H. Chau Nguyen for helpful discussions. This research is supported by the U.S. Air Force Office of Scientific Research (AFOSR) via IOE Grant No. FA9550-19-1-7044 LASCEM, by the European Union’s Horizon Europe programmes HORIZON-CL4-2022-QUANTUM-02-SGA via the project 101113690 (PASQuanS2.1) and HORIZON-CL4-2021-DIGITAL-EMERGING-02-10 under grant agreement No. 101080085 (QCFD), by the Austrian Science Fund (FWF) through the grants SFB BeyondC (Grant No. F7107- N38) and P 32273-N27 (Stand-Alone Project), by the Simons Collaboration on Ultra-Quantum Matter, which is a grant from the Simons Foundation (651440, P.Z.), and by the Institut für Quanteninformation. Innsbruck theory is a member of the NSF Quantum Leap Challenge Institute Q-Sense. TK and BK acknowledge funding from the BMW endowment fund. The computational results presented here have been achieved (in part) using the LEO HPC infrastructure of the University of Innsbruck. All codes and data supporting the findings of this work are available from the corresponding author upon reasonable request.

\end{acknowledgments}

\appendix

\section{Additional constraints}\label{app:additional_constraints}

We consider the equation of motion of a general observable, $O$, not commuting with $H$, as given in Eq.~\eqref{eq:observable}.
Inserting the ansatz $A(\vec{c})$ for $H$ and $\mathcal{D}(\vec{d})$ for the Liouvillian one obtains again a simple matrix equation;
\begin{equation}
\label{eq:learning_equations_BAL}
M^\mathrm{add} \boldsymbol{c} = \boldsymbol{b}(\boldsymbol{d}),
\end{equation}
where
\begin{equation}
\label{eq:constraint_matrix_additional}
M^\mathrm{add}_{ij} = \int_0^{T} \expval*{ -\mathrm{i} [O,h_j] }_{i,t} \mathrm{d}t,
\end{equation}
and
\begin{multline}
\label{eq:constraint_vector_additional}
    b_i(\vec{d}) = \expval*{O}_{i,T} - \expval*{O}_{i,0} \\ - \frac{1}{2} \sum_k d_k \int_0^{T} \expval*{a_k^\dagger [O, a_k] + \mathrm{h.c.}}_{i,t} \, \mathrm{d}t.
\end{multline}
Note, that similar constraints have also been used in Ref.~\cite{Bairey2020} to learn Liouvillians from their steady-states. Compared to the conditions in Eq.~\eqref{eq:learningEq}, the constraints above contain additional integrals in Eq.~\eqref{eq:constraint_matrix_additional} that need to be estimated from experimental data. To obtain the reconstructed parameters we solve the combined system of equations including the additional constraints, i.e.,
\begin{multline}
\label{eq:learning_cost_combined}
\boldsymbol{c}_0^\mathrm{rec}(\xi) = \\
\underset{\boldsymbol{c}}{\mathrm{arg}\,\mathrm{min}} \left[ \min_{\boldsymbol{d}\geq 0} \Big\Vert \begin{pmatrix} M (\boldsymbol{d}) \\ \xi M^\mathrm{add}  \end{pmatrix}
\boldsymbol{c} -
\begin{pmatrix}
0 \\ \xi \boldsymbol{b}(\boldsymbol{d})
\end{pmatrix} \Big\Vert \right],
\end{multline}
where the parameter $\xi$ controls the relative weight between the constraints defined by $M(\vec{d})$ and $M^\mathrm{add}$. In analogy to Eq.~\eqref{eq:learning_cost}, solving Eq.~\eqref{eq:learning_cost_combined} requires a simultaneous minimization over $\boldsymbol{c}$ and $\boldsymbol{d}$, while $\xi$ serves as a 'hyper-parameter' of the optimization. For $\xi=0$  the vector $\boldsymbol{c}_0^\mathrm{rec}(0)$ may be a linear combination of the Hamiltonian and additional, linearly independent, conserved quantities due to the degeneracy of the spectrum of $M(\vec{d})$, as discussed in the main text. Then, by choosing the value of $\xi$ large enough, one removes components of conserved quantities from $\boldsymbol{c}_0^\mathrm{rec}$. In a similar way, one can choose additional constraints such that the solution for the dissipation rates $\boldsymbol{d}$ becomes unique.

Note that the norm of $\boldsymbol{c}_0^\mathrm{rec}$ in Eq.~\eqref{eq:learning_cost_combined} depends on $\xi$ and becomes exact only in the limit $\xi \to \infty$. A finite $\xi$ typically leads to a smaller value for the overall scale of the Hamiltonian. This is because the homogeneous part of Eq.~\eqref{eq:learning_cost_combined} is perfectly solved for $\boldsymbol{c}=0$. Nevertheless, one obtains a $\boldsymbol{c}^\mathrm{rec}_0\propto\boldsymbol{c}^H$. Then, the correct overall scale $s$, defined by $\boldsymbol{c}^H=s\cdot\boldsymbol{c}^\mathrm{rec}_0$, can be determined solely via the additional constraints
\begin{equation}
    \sum_j M^\mathrm{add}_{ij} (s \boldsymbol{c}_0^\mathrm{rec})_j = b_i(\vec{d}^{\rm rec}).
\end{equation}
where $\boldsymbol{d}^\mathrm{rec}$ are the optimal dissipation rates determined from Eq.~\eqref{eq:learning_cost_combined}. Then, averaging over all additional constraints
\begin{equation}
\label{eq:overall_scale}
    s = \frac{1}{p} \sum_{i=1}^p \frac{b_i(\boldsymbol{d}^\mathrm{rec})}{(M^\mathrm{add} \boldsymbol{c}_0^\mathrm{rec})_i}
\end{equation}
yields the correct overall scale.

\section{Effect of shot noise}\label{app:shot-noise}

Here, we discuss in more detail the role of the two smallest singular values of the constraint matrix. As an example we will consider Eq.~\eqref{eq:learningEq}, but similar arguments also apply to Eq.~\eqref{eq:LearningEqEhrenfest}.

In case the ansatz for the dynamics is sufficient, the only source of error is shot noise. Therefore, with a finite measurement budget one obtains a noisy estimate, $\widetilde{M}(\vec{d})=M(\vec{d})+E(\vec{d})$, of the true constraint matrix $M(\vec{d})$, with an additive error matrix $E(\vec{d})$. One can establish the following bound on the perturbed singular value $\lambda_1$
\begin{equation}
    \lambda_1[\widetilde{M}(\vec{d}^\mathrm{rec})]\leq \lambda_1[\widetilde{M}(\vec{d}^\gamma)] \leq \norm{E(\vec{d}^\gamma)},
\end{equation}
where the first inequality holds as $\vec{d}^\mathrm{rec}$ is the minimum of the cost-function in Eq.~\eqref{eq:learning_cost}, and the second inequality follows from Weyl's inequality and the fact that $\lambda_1[M(\vec{d}^\gamma)]=0$ for a sufficient ansatz. Therefore, in early stages of the learning procedure, one expects $\lambda_1$ to decrease $\sim N_{\rm runs}^{-1/2}$ as we increase the number of runs, $N_{\rm runs}$, in the experiment. As the ansatz is sufficient $\lambda_1[\widetilde{M}(\vec{d}^\mathrm{rec})]$ will be zero in the absence of shot noise. However, if an ansatz is insufficient, $\lambda_1$ is strictly bounded away from zero, even in the absence of shot noise. Thus, for $N_{\rm runs}$ sufficiently large, $\lambda_1$ will reach a plateau. Here, the dominant source of error will be systematic errors due to missing terms in our ansatz.

In case of a sufficient ansatz and without degeneracy of the smallest singular values, one can understand the role of the singular value $\lambda_2$. To this end, one considers the angle $\theta=\angle (\vec{c}^H,\vec{c}^{\rm rec})$. It is a well known result in singular subspace perturbation theory, that the stability of a singular vector under perturbation depends on the gap between the corresponding singular value and the remainder of the spectrum, which is known as the Davis-Kahan-Wedin $\sin(\theta)$-theorem~\cite{DavisKahan1970,Wedin1972}. For the case we consider this theorem establishes the following upper bound on the angle $\theta$~\cite{Wedin1972}
\begin{equation}\label{eq:DKW}
    \vert \sin(\theta) \vert \leq\frac{\norm{E(\vec{d}^\mathrm{rec})+R(\Delta \vec{d})}}{\delta},
\end{equation}
where $R(\Delta \vec{d})=\sum_k\Delta d_k M_D^{(k)}$. Here, $\delta=\lambda_2-\lambda_1^{\rm exact}$, and $\lambda_1^{\rm exact}=0$ for a sufficient ansatz. In case the elements of $E$ are i.i.d. Gaussian random variables one can establish average bounds on $\vert \sin(\theta) \vert$~\cite{Li2020}. However, we emphasize that the noise will in general be correlated due to the fact that we perform many of the required measurements in parallel. Moreover, the error matrix $E (\vec{d})$ is experimentally inaccessible. Nevertheless, as we demonstrate in Section~\ref{sec:examples}, the ratio $\lambda_1/\lambda_2$, which can be directly computed from experimental data, serves as a quantity to asses an ansatz.

\section{(Re-)Parameterization}
\label{app:reparametrization}

\begin{figure*}
\subfloat[\label{fig:appendixa}]{\includegraphics[width=0.3\linewidth]{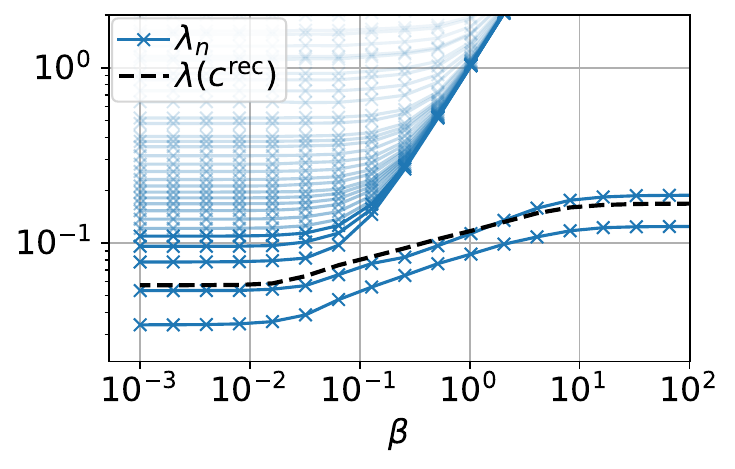}
}
\caption{Singular value spectrum of $M_{G}(\beta)$ defined in Eq.~\eqref{eq:constraints_regularized_beta} as a function of the regularization strength $\beta$ for the model introduced in Section~\ref{sec:ex3_collective} in the main text. Increasing $\beta$ opens a gap between the image of the parametrization matrix $G$ introduced in Appendix~\ref{app:reparametrization} and the rest of the spectrum. Within this subspace, $\lambda(\boldsymbol{c}^\mathrm{rec})$, defined in Eq.~\eqref{eq:lambda_crec}, follows the best approximation to the Hamiltonian of the system (dashed line).
}
\label{fig:appendix}
\end{figure*}

We want to discuss here in more detail the different possibilities to reparametrize an ansatz. In general, there are two possibilities, but we will show that both can be understood in terms of regularization of the cost function in Eq.~\eqref{eq:learning_cost}. Note, that the same holds for Eq.~\eqref{eq:learning_cost_Ehrenfest}.

In the first reparametrization the vector of parameters, $\vec{c}$, is mapped via a \emph{parametrization matrix} $G$ to a new vector of parameters, $\vec{c}_G=G^T\vec{c}$. Here, the matrix $G$ is a $n\times\tilde{n}$ matrix, with $n\geq \tilde{n}$, encoding the dependencies between the $n$ parameters in $\vec{c}$, and the $\tilde{n}$ parameters in $\vec{c}_G$. Moreover, one requires that $G = (\vec{g}_1,\dots ,\vec{g}_{\tilde{n}})$, with orthonormal columns $\langle \vec{g}_i \vert \vec{g}_j \rangle = \delta_{ij}$. One can easily verify that $G^TG=\id_{\tilde{n}\times\tilde{n}}$, and $GG^T=\sum_{i=1}^{\tilde{n}} \ketbra{\vec{g}_i}$, which is a projector onto the support of $G^T$, and thus $G$ is an isometry. In particular cases $G$ can also depend on non-linear parameters, i.e.,
\begin{equation}
\label{eq:parametrization_matrix}
    G = G(\boldsymbol{\alpha}) = (\vec{g}_1(\boldsymbol{\alpha}), \dots, \vec{g}_{\tilde{n}}(\boldsymbol{\alpha})).
\end{equation}
In either case, the new ansatz is given by $A_G(\boldsymbol{c}_G) = A(G \boldsymbol{c}_G)$. This also transforms the constraint matrices via
\begin{equation}
    M^\mathrm{H}_G = M^\mathrm{H}G , \quad M^\mathrm{D}_G(\boldsymbol{d}) = M^\mathrm{D}(\boldsymbol{d})G,
\end{equation}
where the columns of the new constraint matrices are obtained as linear combinations of the columns of the old constraint matrices. The parametrized reconstructed parameters can be obtained as solutions of
\begin{equation}
\label{eq:learning_cost_nonlinear}
\boldsymbol{c}^\mathrm{rec}_G = \underset{\boldsymbol{c}_G, \Vert\boldsymbol{c}_G\Vert=1}{\mathrm{arg}\,\mathrm{min}} \left[ \min_{\boldsymbol{d}\geq 0} \norm{ [M_G^\mathrm{H} + M_G^\mathrm{D}(\boldsymbol{d})] \boldsymbol{c}_G } \right],
\end{equation}
and similarly for Eq.~\eqref{eq:learning_cost_combined}. In the case where $G$ depends on non-linear parameters $\boldsymbol{\alpha}$, the above optimization also includes a minimization over $\boldsymbol{\alpha}$. Numerically, the optimal $\boldsymbol{\alpha}$ can be found similarly to the optimal dissipation rates $\boldsymbol{d}^\mathrm{rec}$ in Eq.~\eqref{eq:learning_cost}, using the DIRECT algorithm in SciPy.

We wish to emphasize that by using this way of parametrizing an ansatz, one obtains reconstructed parameters, where the dependencies, encoded in the matrix $G$, are exactly fulfilled. Examples for such a reparametrization include, for instance, disregarding operators from the operator content of $A(\vec{c})$. This can be understood as a reparametrization, where $(\vec{g}_i)_k=0$, for all $i\in [0, \tilde{n}]$, and $k\in [0,n]$ for which we want to remove the corresponding operator $h_k$ from the operator content of $A(\vec{c})$.

Instead of imposing an exact parametrization on an ansatz, one may only impose it approximately. In practice, this may be very useful as parametrizations are almost never exactly fulfilled, but only to a very good approximation. To this end, one adds a penalty term to the cost function in Eq.~\eqref{eq:learning_cost}, which acts as a regularizing term, giving preference to solutions approximately admitting a certain parametrization. In case of a parametrization $G$, as defined above, the corresponding optimization problem reads
\begin{multline}\label{aeq:regularized}
\boldsymbol{c}^\mathrm{rec}(\beta) = \underset{\boldsymbol{c}, \Vert\boldsymbol{c}\Vert=1}{\mathrm{arg}\,\mathrm{min}} \left[ \min_{\boldsymbol{d}\geq 0} \norm{ [M_\mathrm{H} + M_\mathrm{D}(\boldsymbol{d})] \boldsymbol{c} }\right. \\ \left. +\beta\norm{(\id-GG^T)\vec{c}}\vphantom{\min_{\boldsymbol{d}\geq 0}} \right],
\end{multline}
where $\beta\in[0,\infty)$ is the regularization strength. For $\beta=0$ this corresponds to the original unconstrained (i.e., unparametrized) problem in Eq.~\eqref{eq:learning_cost}. For non-zero $\beta$ the last term adds a penalty, whenever $\vec{c}$  has a component outside of the range of $GG^T$, where the parametrization implied by $G$ is exactly fulfilled. The larger $\beta$, the more $\vec{c}^{\rm rec}$ is constrained by the parametrization $G$. In the limit $\beta\rightarrow\infty$ the parametrization $G$ is fulfilled exactly, and the optimization problem corresponds to the one in Eq.~\eqref{eq:learning_cost_nonlinear}. This can be seen as follows; the objective function in Eq.~\eqref{eq:learning_cost_nonlinear} can be rewritten as $\norm{ [M_\mathrm{H} + M_\mathrm{D}(\boldsymbol{d})]GG^T \boldsymbol{c} }$, where the feasible region consists of all $\vec{c}$ for which $\vec{c}=GG^T\vec{c}$. This in turn can be written as an unconstrained problem in Eq.~\eqref{aeq:regularized}, where $\beta\rightarrow\infty$.
Note, that solving the minimization problem in Eq.~\eqref{aeq:regularized} is equivalent to solving the linear problem
\begin{equation}\label{aeq:regularized_linear}
\boldsymbol{c}^\mathrm{rec}(\beta) = \underset{\boldsymbol{c}, \Vert\boldsymbol{c}\Vert=1}{\mathrm{arg}\,\mathrm{min}}  \left[ \min_{\boldsymbol{d}\geq 0} \left\Vert \begin{pmatrix}  M_\mathrm{H} + M_\mathrm{D}(\boldsymbol{d}) \\ \beta (\id-GG^T) \end{pmatrix} \boldsymbol{c} \right\Vert \right].
\end{equation}

Finally, one notices that given that $GG^T\vec{c}^H=\vec{c}^H$, i.e., the parametrization does not render an ansatz insufficient, the bound in Eq.~\eqref{eq:DKW} can only tighten. To see this, one observes that $\lambda_2[M_G(\vec{d})]\geq \lambda_2[M(\vec{d})]$, and $\norm{E(\vec{d})G}\leq \norm{E(\vec{d})}\norm{G}=\norm{E(\vec{d})}$ for any isometry $G$. Therefore, reducing the number of parameters in general increases the gap $\delta=\lambda_2$ of the constraint matrix $M(\vec{d})$.

\begin{figure*}
    \centering
    \subfloat[\label{fig:comparisonSina}]{
    \includegraphics[width=0.48\columnwidth]{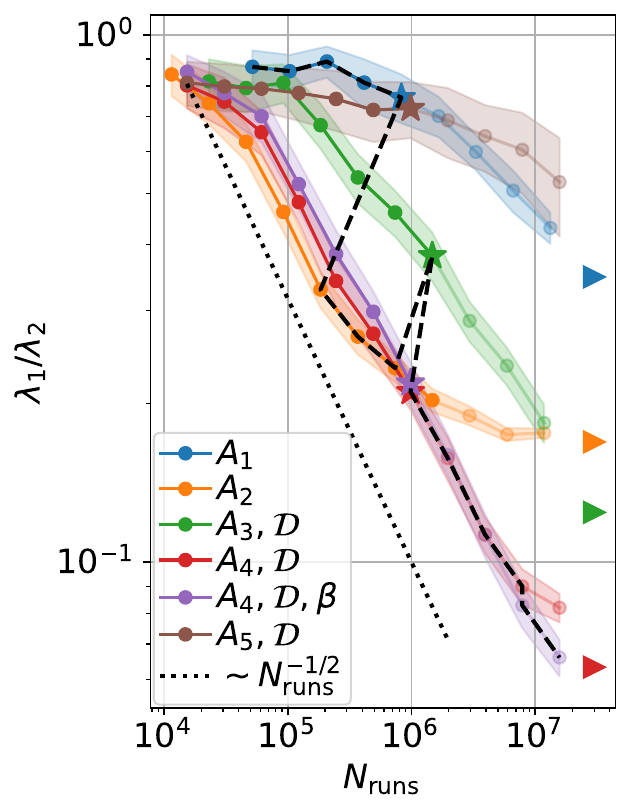}
    }
    \subfloat[\label{fig:comparisonSinb}]{
    \includegraphics[width=0.48\columnwidth]{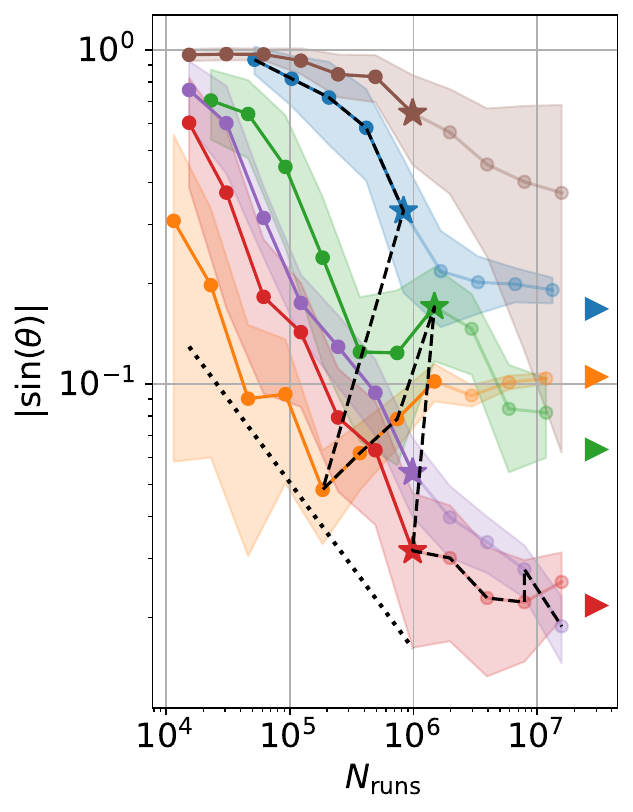}
    }
    \caption{Comparison of $\lambda_1/\lambda_2$ (experimentally accessible), and $\vert \sin(\theta) \vert$ (not experimentally accessible), as a function of the number of runs, $N_{\rm runs}$. The different ansätze are defined in Section~\ref{sec:example1} in the main text. Error bars are computed via bootstrapping as described in Appendix~\ref{app:errorbars}, using $80$ samples. The colored triangles indicate the asymptotic values in the limit $N_{\rm runs}\to\infty$. Both quantities show a similar behaviour which supports using $\lambda_1/\lambda_2$ to asses the quality or learning error of an ansatz.}
    \label{fig:comparisonSin}
\end{figure*}

We now want to study the effect of $\beta$ on the singular value spectrum of the matrix
\begin{equation}
\label{eq:constraints_regularized_beta}
 M_{G}(\beta) = \begin{pmatrix}  M_\mathrm{H} + M_\mathrm{D}(\boldsymbol{d}^\mathrm{rec}) \\ \beta (\id-GG^T) \end{pmatrix},
\end{equation}
for the model system in Section~\ref{sec:ex3_collective} (see Fig.~\ref{fig:appendixa}). 
As expected, in the limit $\beta \to 0$, the spectrum converges to the spectrum of the unparametrized constraint matrix $M(\vec{d}^{\rm rec})$. Here, the gap between the lowest singular values and the rest of the spectrum becomes very small, resulting in an unstable solution.
Then, when increasing $\beta$, all singular values, whose corresponding right-singular vectors are incompatible with the regularization increase with $\beta$, opening a gap to the subspace spanned by the regularization, i.e., the image of $G$, which in the case of the ansatz in Eq.~\eqref{eq:ansatzAXY} is $2$-dimensional. Note, that the Hamiltonian in Eq.~\eqref{eq:model_hamiltonian_FlipFlop} only approximately lies in the image of $G$. Therefore, also the lowest singular values initially increase with $\beta$, until they reach a constant value, that corresponds to an exactly enforced parametrization. Along this path, we can monitor the minimum of the cost function in Eq.~\eqref{aeq:regularized_linear} defined via
\begin{equation}
\label{eq:lambda_crec}
    \lambda(\boldsymbol{c}^\mathrm{rec}) =  \left\Vert \begin{pmatrix}  M_\mathrm{H} + M_\mathrm{D}(\boldsymbol{d}^\mathrm{rec}) \\ \beta (\id-GG^T) \end{pmatrix} \boldsymbol{c}^\mathrm{rec} \right\Vert,
\end{equation}
which follows the singular value that corresponds to the best approximation of the Hamiltonian.

\section{Bootstrapping} \label{app:errorbars}

To obtain error bars from data we use the Bootstrapping method, which we will introduce below.
Assume we are given a single realization $x_1, \dots, x_n$ of a set of independent and identically distributed random variables $X_1, \dots, X_n$ with unknown distribution function. We are interested in estimating the variance of a given function
$T(X_1, \dots, X_n)$. To do so we draw $n$ times with replacement from $x_1, \dots, x_n$, yielding a sample $x_1^\ast, \dots, x_n^\ast$, and then evaluate $t = T(x_1^\ast, \dots, x_n^\ast)$. We repeat this procedure $r$ times obtaining $t^{(1)}, \dots, t^{(r)}$, and then estimate the variance of $T$ from the sample variance of $t^{(1)}, \dots, t^{(r)}$.

In our case the $X_1, \dots, X_n$ are the individual measurements in a given basis for fixed initial state and simulation time. 
One can also think of $X_1, \dots, X_n$ to be individual measurements of a given observable for fixed initial state and simulation time, 
in the case where the measurements of different observables are independent.
Then the quantities of interest $T(X_1, \dots, X_n)$ are, e.g., the learned parameters $\boldsymbol{c}^\mathrm{rec}$, or the ratio of singular values $\lambda_1/\lambda_2$.
The number of samples $r$ is chosen the minimum possible integer, such that the error bars do not significantly change anymore when further increasing $r$.

\section{Comparison between $\lambda_1/\lambda_2$ and $\vert \sin(\theta) \vert$}
In Fig.~\ref{fig:comparisonSin} we provide a comparison between the experimentally measurable quantity $\lambda_1/\lambda_2$, which we use to asses the quality of an ansatz, and the experimentally non-accessible angle between the reconstructed parameters, $\vec{c}^{\rm rec}$, and the Hamiltonian parameters, $\vec{c}^H$, and a function of the number of runs, $N_{\rm runs}$. One observes that both quantities show a very similar behaviour over almost the entire range of $N_{\rm runs}$ considered here. In particular, in the limit $N_{\rm runs}\to\infty$ a larger value of $\lambda_1/\lambda_2$ corresponds to larger $\vert \sin(\theta) \vert$, and vice versa. This supports our idea to use $\lambda_1/\lambda_2$ as an experimentally accessible quantity to asses the quality and learning error of a given ansatz.

\section{Learning from generalized energy conservation and additional constraints}
\label{app:HXYfromEnergyConservation}

\begin{figure*}
\subfloat[\label{fig:example2a}]{
  \includegraphics[width=0.27\linewidth]{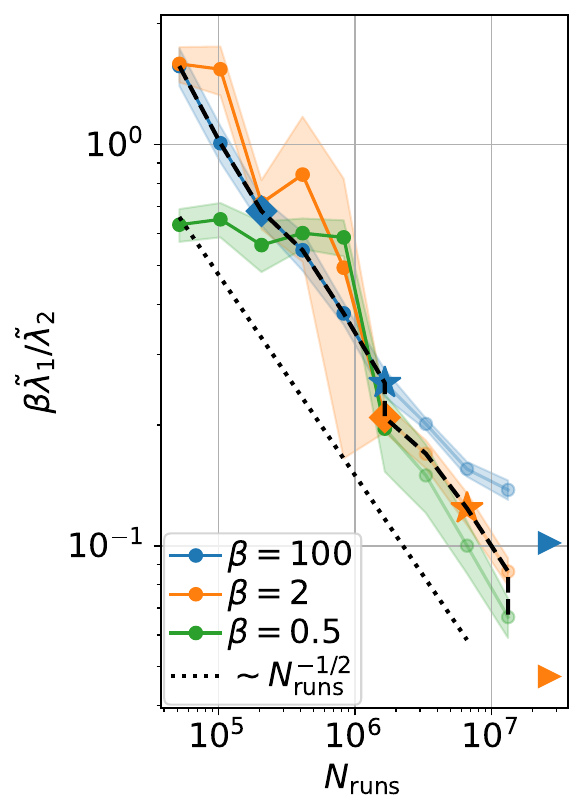}
}\hspace*{1cm}
\subfloat[\label{fig:example2b}]{
  \includegraphics[width=0.61\linewidth]{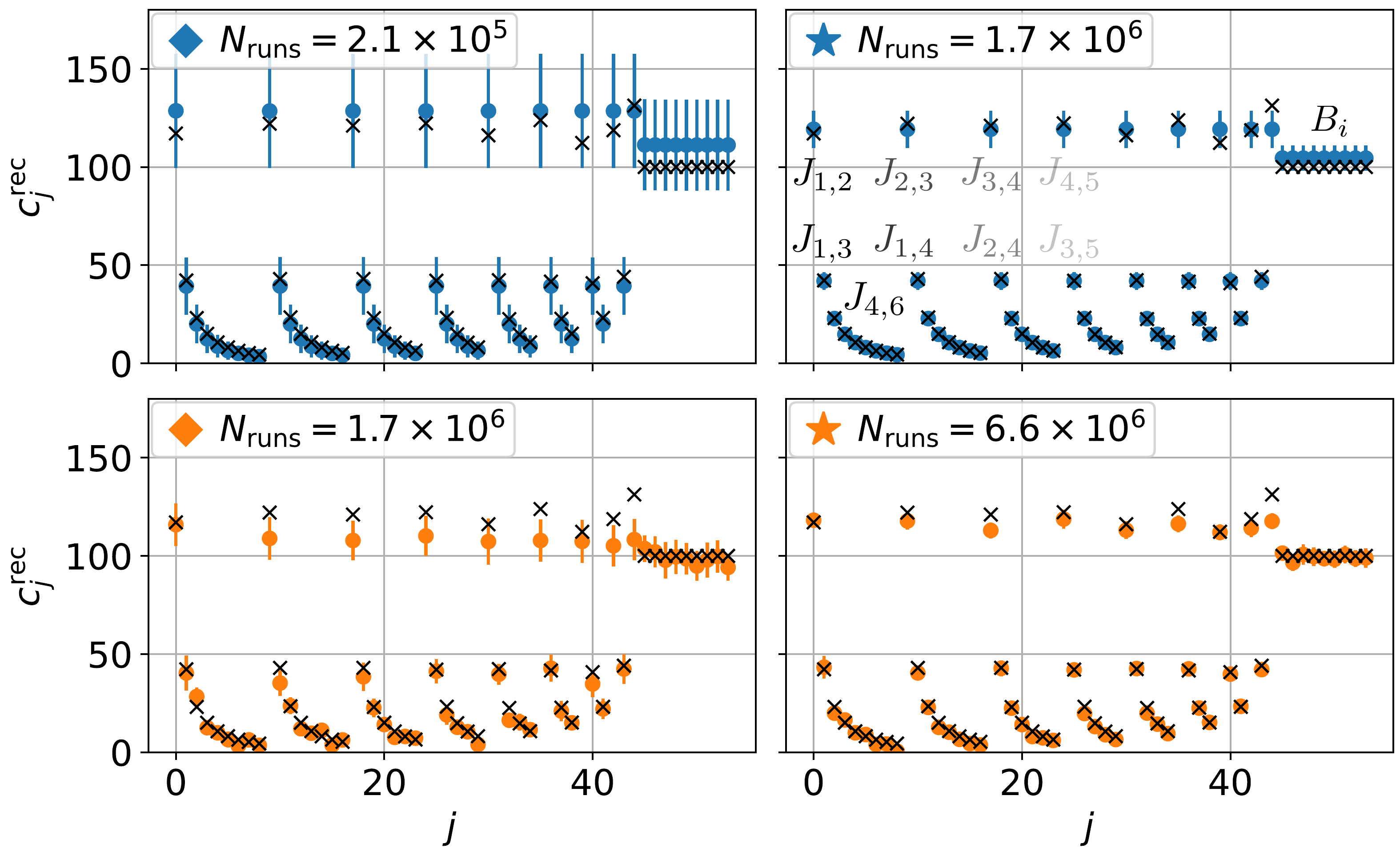}
}
\caption{Hamiltonian learning in presence of collective dissipation. (a) The ratio $\tilde{\lambda}_1/\tilde{\lambda}_2$ after excluding the conserved quantity $Q$ for different regularization parameters $\beta$. The colored triangles describe the asymptotic values at $N_\mathrm{runs}\to\infty$. The dashed line follows the learning procedure described in the main text. (b) Snapshots of the learned Hamiltonian parameters, $c^{\rm rec}_j$, corresponding to the parameters $J_{kl}$ and $B_i$ defined in Eq.~\eqref{eq:model_hamiltonian_FlipFlop}. The required measurement budget is also indicated in the left panel. Black crosses correspond to the parameters of the model Hamiltonian specified below. All error bars are computed via bootstrapping as explained in Appendix~\ref{app:errorbars}, using $10$ samples. ~\eqref{eq:timeTraces} are evaluated using $B_z \Delta_T = 1/128$, i.e., using $128$ time-steps, where at each time we use only a fraction of $1/128$ number of measurements per basis, compared to the quench times $B_z T \in [0,0.5,1]$. Expectation values at $t=0$ are calculated exactly from the initial state. The auxiliary parameter $\xi$ for the additional constraints as introduced in Eq.~\eqref{eq:additionalConstraints} is set to $\xi=1000$.}
\label{fig:example2}
\end{figure*}

In the following we compare the learning based on the generalized Ehrenfest theorem, presented in Section~\ref{sec:ex3_collective} in the main text, to the leaning based on energy conservation. We consider the same model systems comprising the long-range Hamiltonian $H_\mathrm{XY}$ as well as local and collective dissipation, introduced in Section~\ref{sec:ex3_collective} in the main text. 

We will use Eq.~\eqref{eq:ansatzAXY} as an ansatz for the Hamiltonian. As we also want to learn the dissipative processes, we will start with an ansatz for the Liouvillian, comprising single-qubit Lindblad operators that typically appear in the context of trapped-ion experiments, i.e.,
\begin{equation}\label{eq:ansatz_dissipators}
    \mathcal{D}^\mathrm{loc}(\varrho)=\sum_{\eta\in\{+,-,z\}}\frac{d_\eta}{2} \sum_{k=1}^{N} \qty( [\sigma_k^\eta \varrho, \sigma_k^{\eta,\dagger}] + \mathrm{h.c.}).
\end{equation}
Note that this ansatz does not include collective dissipation as given by the off-diagonal elements of $\Gamma^z_{kl}$ in the model Liouvillian. Nevertheless, we will find, that the ansatz in Eq.~\eqref{eq:ansatz_dissipators} is sufficient for learning the Hamiltonian, and we will discuss in detail why this is the case.

Firstly, one notices that the ansatz $A$ contains the total magnetization defined by
\begin{equation} \label{eq:magnetization}
    Q = B \sum_k \sigma^z_{k},
\end{equation}
which commutes with $H_{XY}$ for any $B \in \mathbb{R}$. Therefore, $Q$ is a conserved quantity of $\mathcal{L}_H$. This leads to a degeneracy of the singular value spectrum of the constraint matrix $M(\vec{d})$, i.e., we observe, that $\lambda_1\approx\lambda_2$. In particular, this means that the reconstructed parameters $\vec{c}^{\rm rec}$ will be a linear combination of the parameters of $H_\mathrm{XY}$ and $Q$.
To reconstruct the parameters of the Hamiltonian, including its overall-scale, we need to exclude $Q$, as well as scalar multiples of the Hamiltonian, $\nu\cdot H_\mathrm{XY}$, as possible solutions for $\boldsymbol{c}^\mathrm{rec}$. This is achieved by imposing additional constraints as discussed in Section~\ref{sec:energyConservation}. To this end, we choose a set of operators, $\mathcal{O}=\{ \sigma_1^x,\sigma_1^y, \sigma_1^z\}$, where some operators do not commute with $Q$, while others do not commute with $H_\mathrm{XY}$. Here, one could in principle also choose random operators. This leads to a combined linear optimization as in Eq.~\eqref{eq:additionalConstraints} [see also Appendix~\ref{app:additional_constraints} for the detailed structure of the additional constraints]. 
Moreover, one can define a projected constraint matrix with non-degenerate spectrum $\tilde{\lambda}_1 \leq \tilde{\lambda}_2 \leq \dots$, where $\vec{c}^{\rm rec}\approx \vec{c}^H$ is the right-singular vector corresponding to $\tilde{\lambda}_1$.  Then, $\tilde{\lambda}_1/\tilde{\lambda}_2$ becomes the analog of $\lambda_1/\lambda_2$~\footnote{In the case of two linearly independent conserved quantities our figure of merit $\lambda_1/\lambda_2\approx 1$ independent of $N_{\rm runs}$. Nevertheless, we can define a new matrix $\tilde{M}(\vec{d})$, by projecting the kernel of $M(\vec{d})$, spanned by all vectors $\vec{c}$ that belong to conserved quantities of $H$, onto the vector of reconstructed parameters of the Hamiltonian $\vec{c}^{\rm rec}$. The spectrum $\tilde{\lambda}_1\leq\tilde{\lambda}_2\leq ...$ of $\tilde{M}(\vec{d})$ is gapped, and $\tilde{\lambda}_1/\tilde{\lambda}_2$ becomes the analogue of $\lambda_1/\lambda_2$. For more than two linearly independent conserved quantities one proceeds similarly.}.

In Fig.~\ref{fig:example2a} we monitor $\tilde{\lambda}_1 / \tilde{\lambda}_2$ as a function of $N_\mathrm{runs}$ for the ansatz $A$, and for different values of the regularization parameter $\beta$. One starts with a large value of the regularization parameter, here, $\beta = 100$, which strongly imposes the parametrization of $A$, leading to reconstructed parameters without spatial disorder, see Fig.~\ref{fig:example2b} for $N_\mathrm{runs} = 2.1 \times 10^5$ (blue data). Then further increasing $N_\mathrm{runs}$ reduces the size of the error bars in $\boldsymbol{c}^\mathrm{rec}$, as is also shown in Fig.~\ref{fig:example2b} for $N_\mathrm{runs} = 1.7 \times 10^6$ (blue data). However, an ansatz with $\beta=100$ cannot account for disorder in the coupling terms in $H_{XY}$, which is shown by the plateau in $\tilde{\lambda}_1 / \tilde{\lambda}_2$, here, starting at around $N_\mathrm{runs} \approx 10^7$ (blue line), which, again, is above our measurement budget. Nevertheless, we decrease the regularization parameter $\beta$, which leads to larger error bars, but also allows to learn some of the spatial disorder in $H_\mathrm{XY}$, as is shown in Fig.~\ref{fig:example2b} for $\beta=2$ and $N_\mathrm{runs}=1.7 \times 10^6$ (orange data). Note that already before reaching a plateau for $\beta=2$ in Fig.~\ref{fig:example2a} (orange line) the error bars of $\boldsymbol{c}^\mathrm{rec}$ become very small, which can be seen in Fig.~\ref{fig:example2b} for $\beta=2$ and $N_\mathrm{runs}=6.6 \times 10^6$ (orange data). This suggest to further reduce $\beta$. Ultimately, this process of subsequently reducing $\beta$ and increasing $N_\mathrm{runs}$ allows us to learn $H_\mathrm{XY}$ up to statistical errors, and leads to $\tilde{\lambda}_1 / \tilde{\lambda}_2 \sim N_\mathrm{runs}^{-1/2}$ for the entire range of $N_\mathrm{runs}$. 

\begin{figure*}
    \subfloat[\label{fig:example2_dissa}]{
    \includegraphics[width=0.25\linewidth]{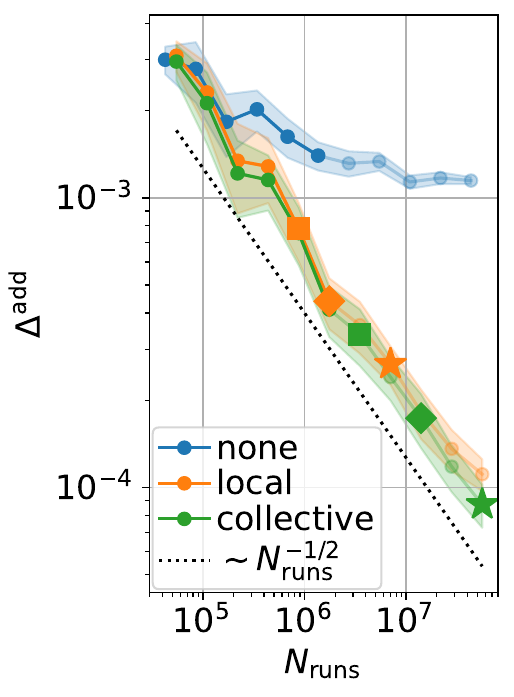}
    }\hspace*{1cm}
    \subfloat[\label{fig:example2_dissb}]{
    \includegraphics[width=0.53\linewidth]{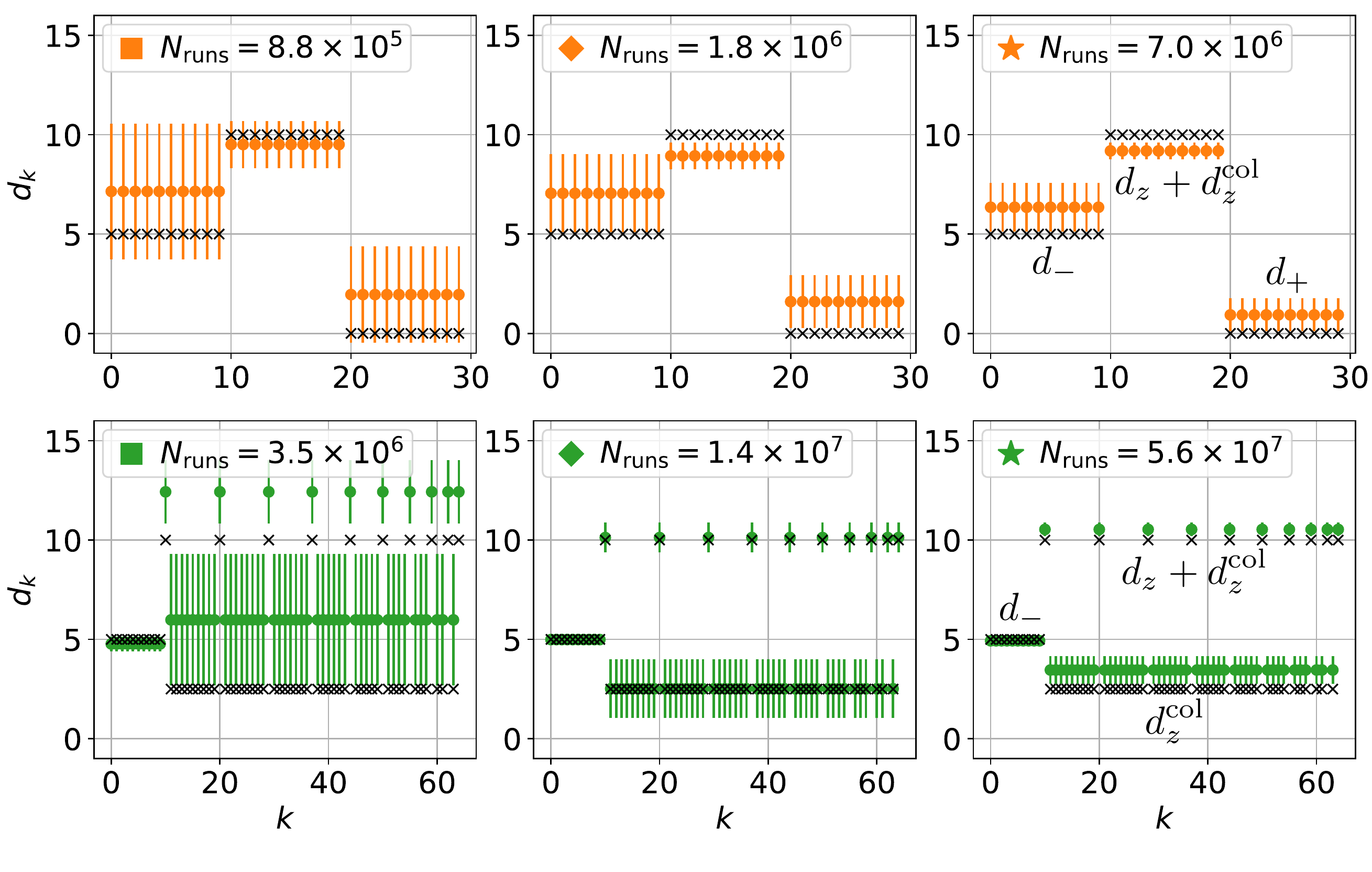}
    }
    \caption{Learning collective dissipation using the additional constraints defined in the main text, and regularization strength $\beta=2$. (a) The error of the additional constraints $\Delta^\mathrm{add}$ as introduced in Eq.~\eqref{aeq:error_signature_BAL} for different ansätze for the dissipation as described the main text. (b) Snapshots of the learned dissipation rates $d_k^\mathrm{rec}$ as indicated by the asterisks in the left panel. Black crosses indicate the true dissipation rates.
    All error bars are computed via Bootstrapping
    as explained in Appendix~\ref{app:errorbars},
    using $10$ samples.
    Integrals for dissipation correction are evaluated as for Fig.~\ref{fig:example2}. We further restrict $d_-, d_z, d_z^\mathrm{col} \geq 0$ and numerically search for the optimal rates within the bounds.}
    \label{fig:example2_diss}
\end{figure*}

So far we have used an ansatz for the Liouvillian which does not include the collective dephasing terms in Eq.~\eqref{eq:collective_dissipation}. Nevertheless, we can learn the Hamiltonian to a high accuracy, which constitutes a useful feature of our approach. To see why this is the case, we evaluate the matrix $M_D(\boldsymbol{d})$ by summing Eq.~\eqref{eq:timeTraces} over all Lindblad operators, yielding
\begin{multline} \label{eq:dissipation_correction_example2}
    [M_D(\boldsymbol{d})]_{\kappa,(i,j)} = \frac{1}{2}\sum_{k,l} d_{kl} [M^{(k,l)}]_{\kappa,(i,j)}  \\ = \frac{1}{2} \int_0^T \sum_{k,l} d_{kl}\expval*{\sigma_k^z[h_{ij},\sigma_l^z]+\mathrm{h.c.}}_{\kappa,t}\d t,
\end{multline}
where $\kappa$ labels input states. Here, we only include the interaction terms, $h_{ij} = \sigma_i^x \sigma_j^x + \sigma_i^y \sigma_j^y$, of the Hamiltonian $H_{XY}$, since the contribution of the collective dissipation in Eq.~\eqref{eq:collective_dissipation} vanishes for the field terms $\sigma_i^z$. Considering the sum over operators appearing in the expectation value in Eq.~\eqref{eq:dissipation_correction_example2}, we can split this sum into
\begin{multline*}
    \sum_{k=1}^N d_{kk} \Big(\sigma_k^z[h_{ij},\sigma_k^z]+ \mathrm{h.c.}\Big) 
    +  \sum_{k\neq l}^N d_{kl} \Big(\sigma_k^z[h_{ij},\sigma_l^z]+ \mathrm{h.c.}\Big) \\
    = -4\sum_{k=1}^N  d_{kk}(\delta_{ki}+\delta_{kj}) h_{ij} +
    4\sum_{k\neq l}^N d_{kl} (\delta_{ki}\delta_{lj}+\delta_{kj}\delta_{li}) h_{ij},
\end{multline*}
which vanishes for $d_{kl}= d_0$. Therefore, the constraint matrix in Eq.~\eqref{eq:learningEq} and hence also the reconstructed parameters of the Hamiltonian are not affected by including collective dissipation into our ansatz. However, this will change if we include additional constraints, as we demonstrate in the following.

\subsection{Learning collective dissipation}

To discover the collective dissipation given by Eq.~\eqref{eq:collective_dissipation} we choose the following ansatz for the Liouvillian
\begin{eqnarray}\label{eq:ansatz_dissipators2}
    \mathcal{D}^\mathrm{col}(\varrho) &&=\frac{d_-}{2} \sum_{k=1}^{N} \qty( [\sigma_k^- \varrho, \sigma_k^{-,\dagger}] + \mathrm{h.c.})  \nonumber \\ && + \sum_{k,l=1}^N (d_z\delta_{kl}+d_z^{\rm col} (1-\delta_{kl})) \mathcal{D}^z_{kl}(\varrho),
\end{eqnarray}
that includes single-qubit spontaneous decay with jump operator $\sigma_k^-$, and single-qubit as well as collective dephasing $\mathcal{D}^z_{kl}$ with operator structure as described in Eq.~\eqref{eq:collective_dissipation}. 
Learning collective dissipation requires the set of constraint operators to also include two-qubit terms. This is because single-qubit operators are not affected by the multi-qubit Lindblad operators in Eq.~\eqref{eq:collective_dissipation}. Therefore, we may choose the following set of constraint operators
\begin{equation*}
\mathcal{O} = \{ \sigma_1^x,\sigma_1^y,\sigma_1^z, \sigma_2^x,\sigma_2^y,\sigma_2^z, \sigma_1^x \sigma_2^x, \sigma_1^x \sigma_2^y, \sigma_1^y \sigma_2^x, \sigma_1^y \sigma_2^y \}.
\end{equation*}
Note that we included here the local operators despite the fact that they cannot help to learn the collective dissipation. This is due to the facts that their expectation values can be obtained from post-processing the data obtained from measuring the two-body operators and that they can be used to better learn the overall scale and the local dissipation rates, as discussed above. Let us emphasize, that the ratio $\lambda_1/\lambda_2$ is not affected by taking collective dissipation in the ansatz into account. Therefore, we instead use
\begin{equation}\label{aeq:error_signature_BAL}
    \Delta^\mathrm{add} \vcentcolon = \Vert M^{\mathrm{add}} \boldsymbol{c}^\mathrm{rec} - \boldsymbol{b}(\boldsymbol{d}^\mathrm{rec}) \Vert ,
\end{equation}
as a measure of how well the additional constraints are fulfilled. In case an ansatz for the Liouvillian is insufficient, not any possible additional constraint can be exactly fulfilled. In such a case, $\Delta^\mathrm{add}$ will be bounded away from zero in the limit $N_{\rm runs}\to\infty$. However, in practice, with only a few additional constraints, e.g., the local Pauli operators discussed above, this must not be the case.

In Fig.~\ref{fig:example2_dissa} we monitor $\Delta^\mathrm{add}$ as a function of $N_\mathrm{runs}$ for the Hamiltonian ansatz in Eq.~\eqref{eq:ansatzAXY} with $\beta=2$, for different ansätze for the Liouvillian. For an ansatz that does not include dissipation (blue line in Fig.~\ref{fig:example2_dissa}) we observe a plateau of $\Delta^\mathrm{add}$ already at very early stages of the learning procedure, here at $N_{\rm runs}\approx 10^5$. On the other hand, we find that the ansätze containing local (orange line) and collective (green line) dissipation lead to similar values of $\Delta^\mathrm{add}$ in early states of the learning procedure. Only above $N_\mathrm{runs} \approx 10^7$ the insufficiency of the local ansatz Eq.~\eqref{eq:ansatz_dissipators} would become evident from considering $\Delta^\mathrm{add}$, which is beyond our available measurement budget. Nevertheless, we can study the 
reconstructed parameters shown in Fig.~\ref{fig:example2_dissb} and their corresponding error bars.
For the ansatz with local dissipation (orange data) we find that the learned rates converge to the exact values, with a good reconstruction achieved at around $N_\mathrm{runs}\approx 10^6$.
Note that here the learned dephasing rate is the sum $d_z + d_z^\mathrm{col}$.
For the ansatz including collective dissipation (green data) we obtain similarly accurate values for $\gamma_-$. However, the local and global dephasing rates, $d_z$ and $d_z^\mathrm{col}$ respectively, have larger error bars. Only around $N_\mathrm{runs}\approx 10^7$ we observe the emergence of non-zero dissipation rates for collective dephasing. This concludes the learning procedure of the Hamiltonian and Liouvillian.

\bibliography{bibliography.bib}

\end{document}